\documentclass[10pt,final,doublecolumn]{IEEEtran}
\usepackage{amsmath}
\usepackage{amssymb}
\usepackage{amsfonts}
\usepackage{latexsym}
\usepackage{graphicx}
\usepackage{bbding}
\usepackage{enumerate}
\usepackage{subeqnarray}
\usepackage{multicol}
\usepackage{color}
\usepackage{setspace}
\usepackage{mathrsfs}
\usepackage{array}
\usepackage{algorithm,algorithmic}
\usepackage{colortbl}
\usepackage{bm}
\IEEEoverridecommandlockouts
\allowdisplaybreaks[3]

\begin{document}
\title{Multiple-Satellite Cooperative Information Communication and Location Sensing in LEO Satellite Constellations}
\author{
Qi Wang, Xiaoming Chen, Qiao Qi, Mili Li, and Wolfgang Gerstacker
\thanks{Qi Wang, Xiaoming Chen, and Mili Li are with the College of Information Science and Electronic Engineering, Zhejiang University, Hangzhou 310027, China (e-mails: wang-qi@zju.edu.cn; chen\underline{~}xiaoming@zju.edu.cn; limili@zju.edu.cn). Qiao Qi is with the School of Information Science and Technology, Hangzhou Normal University, Hangzhou 311121, China (e-mails: qiqiao@hznu.edu.cn). Wolfgang Gerstacker is with the Institute for Digital Communications, Friedrich-Alexander-Universit{\"a}t Erlangen-N{\"u}rnberg, 91058 Erlangen, Germany (e-mails: wolfgang.gerstacker@fau.de). Part of this work has been published by IEEE Wireless Communications and Networking Conference (WCNC), April 2024 \cite{Conf}.
}}\maketitle

\begin{abstract}
Integrated sensing and communication (ISAC) and ubiquitous connectivity are two usage scenarios of sixth generation (6G) networks. In this context, low earth orbit (LEO) satellite constellations, as an important component of 6G networks, is expected to provide ISAC services across the globe. In this paper, we propose a novel dual-function LEO satellite constellation framework that realizes information communication for multiple user equipments (UEs) and location sensing for interested target simultaneously with the same hardware and spectrum. In order to improve both information transmission rate and location sensing accuracy within limited wireless resources under dynamic environment, we design a multiple-satellite cooperative information communication and location sensing algorithm by jointly optimizing communication beamforming and sensing waveform according to the characteristics of LEO satellite constellation. Finally, extensive simulation results are presented to demonstrate the competitive performance of the proposed algorithms.
\end{abstract}


\section{Introduction}
With the rapid development of information technology, there is a growing demand for internet services worldwide. However, traditional terrestrial networks face challenges in providing sufficient coverage to remote areas such as polar and oceanic regions \cite{traditional terrestrial networks}. In this context, satellite communications are widely applied to provide ubiquitous communication services across the globe, and also transform traditional two-dimensional (2D) planar coverage into truly three-dimensional (3D) global spatial coverage. Especially, low earth orbit (LEO) satellites with an orbital altitude less than 2,000 km have attracted significant attention due to the advantages of low transmission latency, propagation attenuation and construction cost, etc. \cite{LEO satellite}. Consequently, LEO satellite constellation communication has been identified as an important component of sixth generation (6G) wireless networks to realize ubiquitous connectivity \cite{satellite industry}. In particular, mega LEO satellite constellation consisting of a large number of LEO satellites works collaboratively to realize seamless coverage and efficient communications on the worldwide scale \cite{LEO Satellite Constellation}. In view of potential prospects of LEO satellite constellations, a series of projects, represented by Starlink, OneWeb, Kuiper and Telesat, are actively promoted and have made significant progress \cite{project}.

More recently, there have been notable advancements in satellite communication technology, bringing us closer to the era of satellite-terrestrial integration \cite{communication technology}. The new-generation LEO satellite systems are leveraging the capabilities of the terrestrial-based telecommunication industry to explore the possibility of relocating base stations (BSs) into space. This innovative approach aims to establish a direct connection between LEO satellites and mobile devices on the ground \cite{direct connection}. In line with this development, AST SpaceMobile achieved a significant milestone in LEO satellite communication technology \cite{AST}. It successfully created a test LEO satellite named BlueWalker 3, achieving impressive data transfer speeds of up to 10 Mbps with standard mobile phones on the ground \cite{AST2}. This successful test serves as a crucial validation of the practicality and viability of direct LEO satellite connectivity. For LEO satellite information communications with direct satellite-terrestrial connections, extensive research has also been conducted in the academic community. For instance, the authors in \cite{LEO communication 1} proposed a pioneering massive multiple-input multiple-output transmission scheme for LEO satellite communications to achieve full frequency reuse. In \cite{LEO communication 2}, the technical problems faced by LEO satellite communications in practical applications were discussed from the perspectives of random access, beam management, and Doppler effect, respectively. Further, for LEO satellite constellation communications, the authors in \cite{LEO constellation communication 1} investigated the constellation coverage problem and proposed a 3D constellation optimization algorithm to achieve seamless global multi-coverage communications. In \cite{LEO constellation communication 2}, the authors performed a communication link design for LEO satellite constellations to provide satellite-terrestrial Internet of Things (IoT) services.

In addition to information communication function, LEO satellites have wide applications in the field of location sensing \cite{LEO satellite location1}. Compared to global location based on medium earth orbit (MEO) and geosynchronous (GEO) satellites located at orbit altitudes of approximately 20,000 km and 35,786 km, respectively, LEO satellite location is suitable for real-time applications and has advantages of precision, accuracy, and system scalability \cite{LEO satellite location2}, \cite{Analysis of LEO}. Furthermore, LEO satellite systems typically possess stronger signal strength and better resistance to interference, facilitating stable location services even in environment where the signal may be obstructed or subjected to interference. There are numerous studies focusing on LEO satellite location as well. For example, in \cite{satellite localization 1}, a novel round trip time-based method was proposed for remote nodes localization by using a single LEO satellite. The authors in \cite{satellite localization 2} summarized traditional satellite-based passive localization techniques and proposed a time difference of arrival-assisted direct position determination method for LEO satellite networks. Moreover, some preliminary studies have been conducted in both academia and industry on location sensing in multiple-satellite or LEO satellite constellation scenarios. In \cite{R1_1}, the authors utilized the Successive Interference Cancellation (SIC) algorithm to separate and reconstruct the reference signals, proposing a method based on detection of weak echo signals from multi-satellite GNSS for passive sensing. The authors in \cite{R1_2} achieved passive location sensing of moving targets by estimating the time difference of arrival (TDOA) and frequency difference of arrival (FDOA) from multiple satellites. Meanwhile, the authors in \cite{LEO satellite location2} presented a theoretical model of LEO global location constellation based on the walker configuration, which satisfies the quadruple global coverage constraints. In industry, SpaceX launched a starshield project for global sensing by making use of its LEO satellite constellation \cite{starshield}.

As 6G wireless networks develop, a series of new applications are emerging, which put forward higher requirements for information communication and location sensing \cite{ISAC 1}. In this context, it is an unstoppable trend to integrate information communication and location sensing into a single system to improve the utilization efficiency of wireless and hardware resources \cite{ISAC 2}. In fact, the concept of integrated sensing and communication has gained significant attention and sparked widespread discussion in terrestrial networks. For instance, some key applications and state-of-the-art approaches to terrestrial integrated sensing and communication were provided in \cite{ground ISAC 1}, along with a discussion of the performance trade-off between the two functions from the signal processing perspective. The authors in \cite{ground ISAC 2} studied a transmit beamforming scheme for a downlink integrated sensing and communication system by maximizing radar sensing performance while meeting communication quality of service requirements. Yet, in non-terrestrial networks, the limited payload processing capacities of traditional LEO satellites result in the separation of information communication and location sensing at different satellites. Fortunately, the enhanced payload processing capabilities of new-generation LEO satellites make it possible to simultaneously conduct multiple functions on a single hardware platform for non-terrestrial networks \cite{satellite ISAC 0}. For instance, the authors in \cite{satellite ISAC 1} investigated a target localization error minimization problem in a radar-communication satellite system, where a dual-function satellite simultaneously served multiple communications users and detected a moving target. In \cite{satellite ISAC 2}, a multi-objective optimization problem in a downlink LEO satellite system was formulated to achieve an overall performance improvement of both communication and target sensing by optimizing the beamforming.

Nevertheless, above studies lack consideration of the megatrend toward multiple-satellite collaboration, focusing only on individual LEO satellite. Driven by this, we aim to exploit the potential of LEO satellite constellations in information communication and location sensing to advance the realization of 6G air-space-ground integration. On the one hand, since the high-speed inter-satellite optical communication links can provide high-capacity data exchange, we can leverage collaborative transmission among multiple LEO satellites to transform the interference signals between different satellites into useful signals, thereby enhancing the quality of received signals at the ground. On the other hand, we can achieve more accurate and reliable location estimation with direct localization method by using received reflecting signals from multiple satellites. However, achieving high-rate information communication and high-precision location sensing simultaneously for LEO satellite constellations is not a trivial task. Especially within limited wireless resources and under high dynamic environment, the inter-satellite interference and the inter-functional interference between information communication and location sensing will greatly restrict the performance of the LEO satellite constellations. Motivated by this, we take advantage of the unique strengths offered by LEO satellite constellations to achieve the comprehensive integrated information communication and location sensing via jointly designing multiple-satellite communication beamforming and sensing waveform. The contributions of this paper are as follows.

\begin{enumerate}
\item We provide a dual-function LEO satellite constellation framework to integrate information communication and location sensing with the same hardware and spectrum.

\item We derive a closed-form expression for the Cramer-Rao bound (CRB) as location sensing metric by using a particle swarm optimization (PSO)-based direct location sensing algorithm, and analyze the impact of key factors on the system performance.

\item we propose a multiple-satellite cooperative information communication and location sensing design by jointly optimizing communication beamforming and sensing waveform to improve location sensing accuracy while maintaining an efficient information transmission rate for LEO satellite constellations.

\end{enumerate}

The subsequent sections are structured as follows: In Section II, we present the system model for dual-function LEO satellite constellation. Section III discusses a design of multiple-satellite cooperative information communication and location sensing. Simulation results are provided in Section IV to verify the efficacy of the proposed algorithms. Finally, Section V presents the concluding remarks of this paper.

\emph{Notations}: Ordinary letters, bold lowercase letters and bold uppercase letters denote scalars, vectors and matrices, respectively. $(\cdot)^T$, $(\cdot)^H$, $(\cdot)^{-1}$, $\text{Rank}(\cdot)$ and $\text{tr}(\cdot)$ indicate the transpose, conjugate transpose, inverse, rank and trace of a matrix, respectively. $|\cdot|$ represents the absolute value of a scalar, $\|\cdot\|$ represents the 2-norm of a vector, $[\mathbf{x}]_i$ represents the $i$-th element of the vector $\mathbf{x}$, $[\mathbf{X}]_{i,j}$ represents the element in the $i$-th row and $j$-th column of matrix $\mathbf{X}$, $\text{Diag}(\cdot)$ indicates the generation of a diagonal matrix, $\mathbf{X}\succeq\mathbf{0}$ implies that matrix $\mathbf{X}$ is positive semi-definite, $\left[ {{\bf{X}},{\bf{Y}}} \right]$ means horizontal concatenation of matrices, and $\left[ {{\bf{X}};{\bf{Y}}} \right]$ means vertical concatenation of matrices. ${{\mathbb{C}}^{a\times b}}$ and ${{\mathbb{R}}^{a\times b}}$ denote the sets of $\emph{a}$ $\times$ $\emph{b}$ dimensional complex and real matrixes, respectively. $\odot$ denotes the Hadamard product and $\otimes$ stands for the Kronecker product. $J_1(\cdot)$ and $J_3(\cdot)$ represent the first-order and third-order Bessel functions, respectively.

\section{System Model}
\begin{figure} \centering
\includegraphics [width=0.50\textwidth] {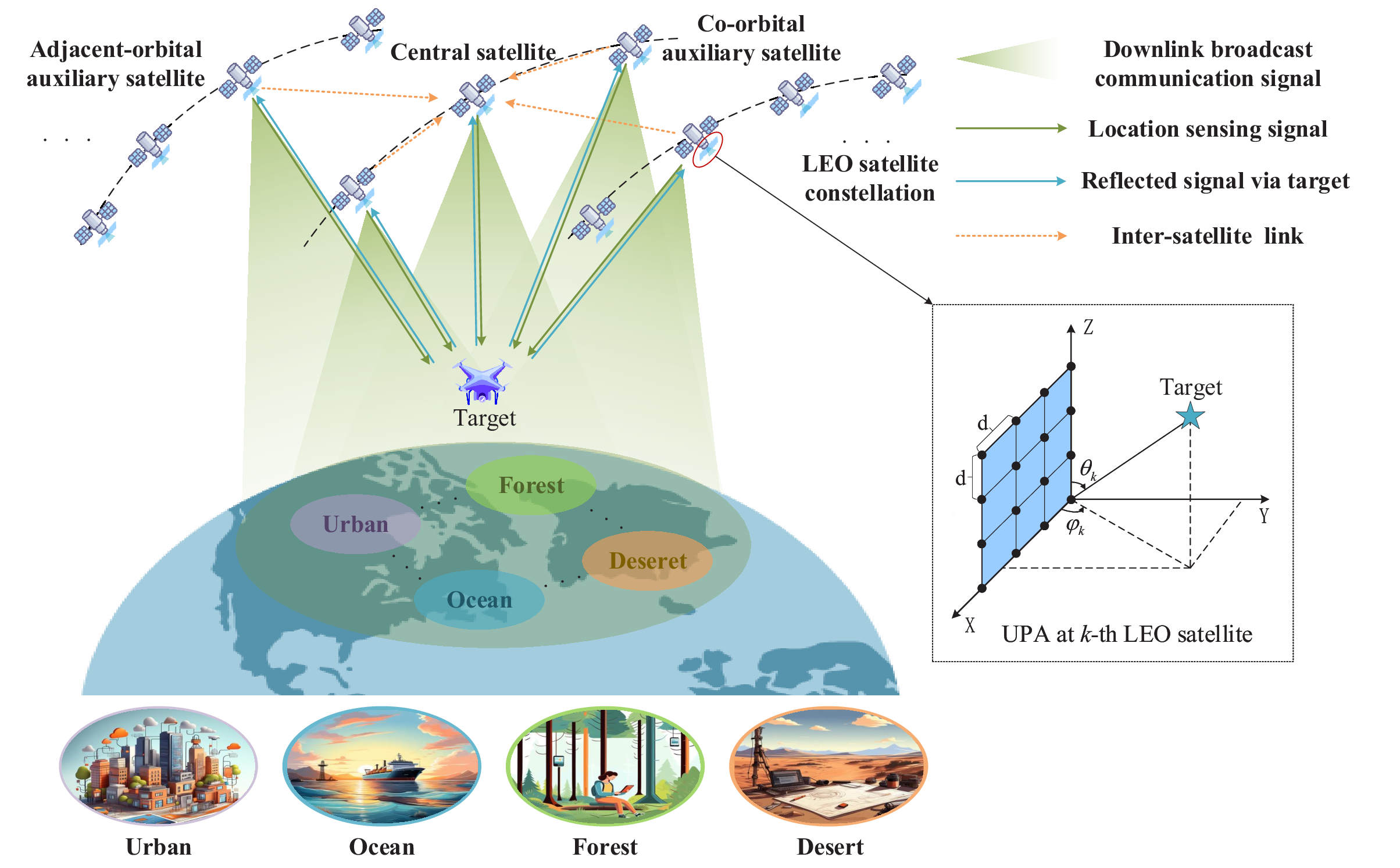}
\caption {System model for dual-function LEO satellite constellation.}
\label{Fig1}
\end{figure}
\begin{figure} \centering
\includegraphics [width=0.4\textwidth] {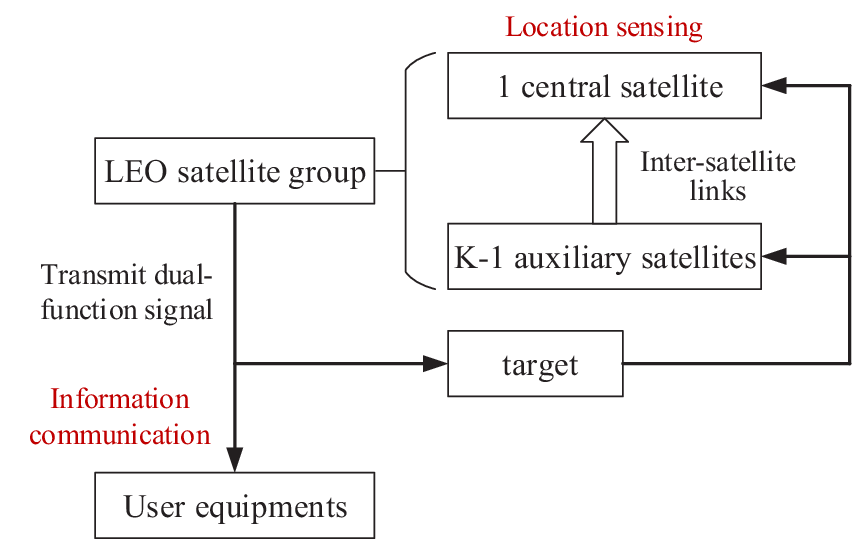}
\caption {System flowchart.}
\label{System flowchart.}
\end{figure}

We consider a Walker Delta configuration-based LEO satellite constellation with orbital altitude $H^{\#}$, $P^{\#}$ orbital planes, $N^{\#}$ LEO satellites per orbital plane, orbital inclination $I^{\#}$, and phase factor $F^{\#}$ \footnote{In Walker Delta constellation, all satellites have the same orbit altitude and are evenly distributed on circular orbits with the same inclination angle. Compared to other configurations, e.g., Walker Star \cite{Walker star} and Rosette \cite{Rosette} constellations, the satellite constellation based on the Walker Delta configuration offers unified global coverage, effective satellite utilization, reduced inter-satellite interference, high scalability and simplicity in design and implementation, and has been well-studied and validated. Thus, the Walker Delta constellation is widely applied for satellite communications, remote sensing, and global navigation systems \cite{Walker}.}. Each LEO satellite is equipped with a uniform planar array (UPA) with $N$ full-duplex antennas. As shown in Fig. 1, a central satellite provides information communication services for $M$ single-antenna ground user equipments (UEs) within the coverage area and location sensing service for an interested target simultaneously with the collaboration of $K-1$ neighboring auxiliary satellites in the same orbital plane and the adjacent orbital planes. These auxiliary satellites are connected with the central satellite via high-capacity inter-satellite links to form a serving group.
As is depicted in Fig. \ref{System flowchart.}, each LEO satellite in this group transmits a dual-function communication and sensing signal over the same spectrum. For information communication, all UEs receive and decode the dual-function signal transmitted over the satellite-terrestrial channel. For location sensing, all LEO satellites in the group receive the dual-function signal reflected by the interested target, and then transfer it to the central satellite through inter-satellite links to further perform the target location estimation.

Specifically, the $k$-th LEO satellite broadcasts the dual-function signal at the time index $t$ as
\begin{equation}\label{broadcast signals}
{{\bf{x}}_k}(t)= \sum\limits_{m = 1}^M {{{\bf{w}}_{k,m}}{s_m}}(t) + {{\bf{r}}_k}(t),
\end{equation}
where ${{\bf{w}}_{k,m}}\in{\mathbb{C}^{N \times 1}}$ denotes the transmit communication beamforming vector at the $k$-th LEO satellite for delivering the zero-mean unit power communication data symbol $s_m$ of the $m$-th UE, and ${{\bf{r}}_k} \in {\mathbb{C}^{N \times 1}}$ denotes the location sensing signal waveform specially designed at the $k$-th LEO satellite to enable full degrees of freedom for target sensing \cite{ground ISAC 2}. Herein, $\left\{ {{s_m}} \right\}_{m = 1}^M$ are assumed to be statistically independent random symbols and are also independent of ${{\bf{r}}_k}$. Thus, the transmit power of the $k$-th LEO satellite is given by
\begin{equation}\label{power}
P_k = \sum\limits_{m = 1}^M {{{\left\| {{{\bf{w}}_{k,m}}} \right\|}^2}}  + {{{\left\| {{{\bf{r}}_{k}}} \right\|}^2}}.
\end{equation}
In the following, we introduce the multiple-satellite cooperative information communication and location sensing models, respectively.

\subsection{Cooperative Information Communication Model}
For information communication, the dual-function signals are transmitted to UEs via downlink satellite-terrestrial channels.
Let vector $\mathbf{h}_{k,m}\in{\mathbb{C}^{N \times 1}}$ denote the downlink satellite-terrestrial channel in Ka band (26.5-40 GHz) between the $k$-th LEO satellite and the $m$-th UE, which can be expressed as \cite{Ying}
\begin{equation}\label{channel model}
{{\bf{h}}_{k,m}} = {{\bf{g}}_{k,m}} \odot \left( {\sqrt {\frac{{{\lambda _k}}}{{{\lambda _k} + 1}}} {\bf{h}}_{k,m}^{{\rm{LOS}}} + \sqrt {\frac{1}{{{\lambda _k} + 1}}} {\bf{h}}_{k,m}^{{\rm{NLOS}}}} \right),
\end{equation}
where $\lambda _k$ represents Rician factor, and ${\bf{h}}_{k,m}^{{\rm{LOS}}}$ and ${\bf{h}}_{k,m}^{{\rm{NLOS}}}$ are the line of sight (LOS) and non-line of sight (NLOS) components of the Rician satellite-terrestrial channel, respectively. Furthermore, ${{\bf{g}}_{k,m}}\in{\mathbb{C}^{N \times 1}}$ represents the channel gain and is expressed as \cite{satellite-terrestrial channel}
\begin{equation}\label{channel gain}
{{\bf{g}}_{k,m}} = \sqrt {{{\left( {\frac{c}{{4\pi f{d_{k,m}}}}} \right)}^2}\frac{{{G_{m}}}}{{\kappa BT}}}  \cdot {{\bm{\chi}}_{k,m}} \odot {\bf{b}}_{k,m}^{\frac{1}{2}},
\end{equation}
where $c$, $f$, $\kappa$, $B$ and $T$ are the speed of light, signal frequency, Boltzmann constant, channel bandwidth and noise temperature, respectively. $d_{k,m}$ denotes the distance between the $k$-th LEO satellite and the $m$-th UE, and $G_{m}$ stands for the receive antenna gain at the $m$-th UE. In addition, ${{\bm{\chi}}_{k,m}}= {\xi ^{\frac{1}{2}}}{e^{ - j{{\bm{\psi }}_{k,n}}}}$ denotes the rain attenuation vector, which is the main component of atmospheric attenuation for high-frequency band links. The term ${\xi ^{\frac{1}{2}}}$ is the rain attenuation gain in dB with a logarithmic normal distribution $\ln \left( {{\xi ^{{1}/{2}}}} \right)\sim\mathcal{C}\mathcal{N}\left( {{\mu _r},\sigma _r^2} \right)$, and ${{\bm{\psi }}_{k,n}}$ is a phase vector \cite{rain attenuation}. Moreover, ${{\bf{b}}_{k,m}}$ is the $N$-dimensional satellite antenna gain, whose element is given by \cite{beam gain}
\begin{equation}\label{satellite antenna gain}
[{{\bf{b}}_{k,m}}]_i = {b_{k}^{\max}}{\left( {\frac{{{J_1}\left( {{u_i}} \right)}}{{2{u_i}}} + 36\frac{{{J_3}\left( {{u_i}} \right)}}{{{u_i}^3}}} \right)^3},
\end{equation}
where ${u_i} = 2.071\left( {{{\sin \left( {{\varepsilon _{i,k,m}}} \right)} \mathord{\left/{\vphantom {{\sin \left( {{\varepsilon _{i,k,m}}} \right)} {\sin \left( {\varepsilon _k^{3dB}} \right)}}} \right.\kern-\nulldelimiterspace} {\sin \left( {\varepsilon _k^{3dB}} \right)}}} \right)$ with ${\varepsilon _{i,k,m}}$ being the angle between the $i$-th antenna of the $k$-th LEO satellite and the $m$-th UE, and $\varepsilon _k^{3dB}$ and ${b_{k}^{\max}}$ are the 3-dB angle and maximum antenna gain of the $k$-th LEO satellite, respectively. In this context, the received signal at the $i$-th UE can be written as
\begin{align}\label{UE received signal}
{y_i}\left( t \right)&= \sum\limits_{k = 1}^K {{\bf{h}}_{k,i}^{\rm H}{{\bf{x}}_k}\left( {t - {\tau _{k,i}}} \right)} {e^{j2\pi {v_{k,i}}t}} + {n_i}\left( t \right)\notag\\
&= \underbrace{\sum\limits_{k = 1}^K {{\bf{h}}_{k,i}^{\rm H}} {{\bf{w}}_{k,i}}{s_i}\left( {t - {\tau _{k,i}}} \right){e^{j2\pi {v_{k,i}}t}}}_{\text{desired signal}} \notag\\
&\ \ \ \ + \underbrace{\sum\limits_{k = 1}^K {\sum\limits_{m \ne i}^M {{\bf{h}}_{k,i}^{\rm H}{{\bf{w}}_{k,m}}{s_m}\left( {t - {\tau _{k,i}}} \right){e^{j2\pi {v_{k,i}}t}}} }}_{\text{inter-UE interference}}  \notag\\
&\ \ \ \ \underbrace{+ \sum\limits_{k = 1}^K {\sum\limits_{m = 1}^M {{\bf{h}}_{k,i}^{\rm H}{{\bf{r}}_k}\left( {t - {\tau _{k,i}}} \right){e^{j2\pi {v_{k,i}}t}}} }}_{\text{sensing interference}}  + \underbrace{{n_i}\left( t \right)}_{\text{noise}},
\end{align}
where $\forall i \in \{1,\ldots,M\}$, ${n_i}$ denotes the additive white Gaussian noise (AWGN) with variance $\sigma _i^2$ received at the $i$-th UE, and ${\tau _{k,i}}$ and ${v_{k,i}}$ are the propagation delay and Doppler shift associated with the channel from the $k$-th LEO satellite to the $i$-th UE caused by long-range transmission and high-speed mobility, respectively.
To improve the accuracy and reliability of signal decoding, joint delay synchronization and Doppler compensation techniques are employed on the receiver at each UE \cite{Doppler}. Hence, by using Doppler shift estimation and compensation methods such as geometric analysis \cite{Geometric Analysis}, Kalman filter \cite{Kalman Filter} and maximum likelihood doppler estimation \cite{MLE Doppler}, it is reasonably assumed that the effect of Doppler shift can be mitigated.

For ease of presentation, we omit the time index and define cooperative communication beamforming ${\tilde{{\bf{w}}}_m} = \left[ {{{\bf{w}}_{1,m}}; \cdots ;{{\bf{w}}_{K,m}}} \right] \in{\mathbb{C}^{NK \times 1}}$, cooperative sensing waveform ${\tilde{\bf{r}}} = \left[ {{{\bf{r}}_1}; \cdots ;{{\bf{r}}_K}} \right] \in{\mathbb{C}^{NK \times 1}}$ and multiple-satellite equivalent channel ${\tilde{{\bf{h}}}_m} = \left[ {{{\bf{h}}_{1,m}}; \cdots ;{{\bf{h}}_{K,m}}} \right] \in{\mathbb{C}^{NK \times 1}}$. As a result, the received signal at the $i$-th UE as described in equation (\ref{UE received signal}) can be reformulated as
\begin{equation}
{y_i} = {\tilde{\bf{h}}}_i^{\rm H}{\tilde{{\bf{w}}}_i}{s_i} + \sum\limits_{m \ne i}^M {\tilde{{\bf{h}}}_i^{\rm H}{\tilde{{\bf{w}}}_m}{s_m}}  + {\tilde{\bf{h}}}_{\rm{i}}^{\rm{{\rm H}}}{\tilde{\bf{r}}} + {n_i}.
\end{equation}

To evaluate the quality of information communication, a commonly used date transmission rate is adopted as the communication performance metric. Specifically, the achievable transmission rate for the $i$-th UE is given by
\begin{equation}\label{R_i}
{R_i} = {\log _2}\left( {1 + \frac{{{{\left| {\tilde{{\bf{w}}}_i^{\rm H}{\tilde{{\bf{h}}}_i}} \right|}^2}}}{{\sum\limits_{m \ne i}^M {{{\left| {\tilde{{\bf{w}}}_m^{\rm H}{\tilde{{\bf{h}}}_i}} \right|}^2}}  + {{\left| {{\tilde{{\bf{r}}}^{\rm H}}{\tilde{{\bf{h}}}_i}} \right|}^2} + \sigma _i^2}}} \right).
\end{equation}
It is observed from equation (\ref{R_i}) that the cooperative communication beamforming $\tilde{{\bf{w}}}_m$ and cooperative sensing waveform $\tilde{{\bf{r}}}$ are the key factors affecting the cooperative information communication performance. As a result, it is likely to improve the communication performance by selecting appropriate $\tilde{{\bf{w}}}_m$ and $\tilde{{\bf{r}}}$.

\subsection{Cooperative Location Sensing Model}
For location sensing, the dual-function signals are reflected by the interested target and then are received by all LEO satellites. Let ${{\bf{q}}_k} = \left( {q_k^x, q_k^y, q_k^z} \right)$ and ${\bf{p}} = \left( {{p^x}, {p^y}, {p^z}} \right)$ represent the location vectors of the $k$-th LEO satellite and the target in Cartesian earth-centered earth-fixed (ECEF) coordinate system, respectively. Therefore, the elevation angle ${\theta _k}\in \left[ {0,\pi } \right]$ and the azimuth angle $\varphi _k \in \left[ { - \pi ,\pi } \right]$ of the target with respect to the $k$-th LEO satellite can be calculated based on geometric locations as
\begin{equation}\label{theta}
{\theta _k} = \arctan \frac{{\sqrt {{{\left( {{p^x} - q_k^x} \right)}^2} + {{\left( {{p^y} - q_k^y} \right)}^2}} }}{{{p^z} - q_k^z}} + \pi  \cdot \mathbb{T}\left[ {{p^z} < q_k^z} \right]
\end{equation}
and
\begin{equation}\label{varphi}
{\varphi _k} = \left\{ \begin{array}{l}
\arctan \frac{{{p^y} - q_k^y}}{{{p^x} - q_k^x}} + \pi  \cdot \mathbb{T}\left[ {{p^x} < q_k^x,{p^y} > q_k^x} \right]\\
\arctan \frac{{{p^y} - q_k^y}}{{{p^x} - q_k^x}} - \pi  \cdot \mathbb{T}\left[ {{p^x} < q_k^x,{p^y} < q_k^x} \right]
\end{array} \right.,
\end{equation}
where $\mathbb{T}\left[\cdot\right]$ is a logical operator representing that the value is 1 if $\left[\cdot\right]$ is true, otherwise the value is 0. Similarly, after performing delay synchronization and Doppler compensation, the received signal at the $u$-th LEO satellite can be expressed as
\begin{equation}
{{\bf{y}}_u} = \sum\limits_{k = 1}^K {\alpha {\beta _{k,u}}{{\bf{a}}_r}\left( {{\theta _u},{\varphi _u}} \right){\bf{a}}_t^{\rm H}\left( {{\theta _k},{\varphi _k}} \right){{\bf{x}}_k}}  + {{\bf{n}}_u},
\end{equation}
where $\forall u\in\{1,\ldots,K\}$, $\alpha$ denotes the reflection coefficient of the target, ${\beta_{k,u}}$ stands for the sensing channel gain from the $k$-th to the $u$-th LEO satellite through target reflection, and ${{\bf{n}}_u}$ denotes the noise vector at the $u$-th LEO satellite. Herein, we assume that clutters are randomly distributed in the environment, treating clutter reflected interference with random phases and magnitudes as a special type of noise absorbed into Gaussian noise \cite{CRB ref2}. In addition, ${{\bf{a}}_r}\left( {{\theta _u},{\varphi _u}} \right) \in {\mathbb{C}^{N \times 1}}$ and ${\bf{a}}_t\left( {{\theta _k},{\varphi _k}} \right) \in {\mathbb{C}^{N \times 1}}$ represent the receive and transmit steering vectors, respectively. Specifically, ${{\bf{a}}_r}\left( {{\theta _u},{\varphi _u}} \right)$ and ${\bf{a}}_t\left( {{\theta _k},{\varphi _k}} \right)$ have the same structure and can be expressed as
\begin{align}
&{\bf{a}}({{\theta _k},{\varphi _k}})=\bigg(\frac{1}{{\sqrt {{N_x}} }}\Big[ 1; \cdots ;{e^{j\frac{{2\pi }}{\lambda }d( {{i_x}\cos {\varphi _k}\sin {\theta _k}} )}}; \cdots ;\notag\\
&{e^{j\frac{{2\pi }}{\lambda }d( {( {{N_x} - 1})\cos {\varphi _k}\sin {\theta _k}} )}} \Big]\bigg)\otimes \bigg(\frac{1}{{\sqrt {{N_z}} }}\Big[ 1; \cdots ;{e^{j\frac{{2\pi }}{\lambda }d( {{i_z}\cos {\theta _k}})}}; \notag\\
&\ \ \ \ \ \ \ \ \ \ \ \ \ \ \ \ \ \ \   \cdots ;{e^{j\frac{{2\pi }}{\lambda }d( {( {{N_z} - 1})\cos {\theta _k}} )}} \Big]\bigg)\notag\\
&=\frac{1}{{\sqrt {{N}} }}\left[ \begin{array}{l}
1; \cdots ;{e^{j\frac{{2\pi }}{\lambda }d\left( {{i_x}\cos {\varphi _k}\sin {\theta _k} + {i_z}\cos {\theta _k}} \right)}}; \cdots \\
;{e^{j\frac{{2\pi }}{\lambda }d\left( {\left( {{N_x} - 1} \right)\cos {\varphi _k}\sin {\theta _k} + \left( {{N_z} - 1} \right)\cos {\theta _k}} \right)}}
\end{array} \right],
\end{align}
where $N={N_x}\times{N_z}$ with $N_x$ and $N_z$ being the number of antennas on the x-axis and z-axis of the UPA, respectively. Furthermore, $\lambda$ is the signal wavelength, $d$ is the adjacent antenna spacing, and ${i_x} \in \left\{ {1, \cdots ,{N_x-1}} \right\}$ and ${i_z} \in \left\{ {1, \cdots ,{N_z-1}} \right\}$ are the antenna indexes. For ease of expression, we define ${{\bf{\tilde{a}}}_t}\left( {{\bm{\theta }},{\bm{\varphi }}} \right) = \left[ {{\bf{a}}_t^{\rm H}\left( {{\theta _1},{\varphi _1}} \right), \cdots ,{\bf{a}}_t^{\rm H}\left( {{\theta _K},{\varphi _K}} \right)} \right] \in {\mathbb{C}^{1 \times NK}}$ with ${\bm{\theta }} = \left[ {{\theta _1}; \cdots ;{\theta _K}} \right]$ and ${\bm{\varphi }} = \left[ {{\varphi _1}; \cdots ;{\varphi _K}} \right]$,
${\bf{B}}\left( {{{\bm{\beta }}_u}} \right) = {\rm{Diag}}\left( {{{\bm{\beta }}_u}} \right) \otimes {{\bf{I}}_N} \in {\mathbb{C}^{NK \times NK}}$ with ${{\bm{\beta }}_u} = \left[ {{\beta _{1,u}}; \cdots ;{\beta _{K,u}}} \right]$,
${\bf{V}} = \left[ {{{\bf{V}}_1}; \cdots ;{{\bf{V}}_K}} \right] \in {\mathbb{C}^{NK \times M}}$ with ${{\bf{V}}_k} = [{{\bf{w}}_{k,1}}, \cdots ,{{\bf{w}}_{k,M}}] \in {\mathbb{C}^{N \times M}}$, and ${\bf{s}} = \left[ {{s_1}; \cdots ;{s_M}} \right]$.
Hence, the received signal at the $u$-th LEO satellite is reformulated as
\begin{equation}
{{\bf{y}}_u} = \alpha {{\bf{a}}_r}\left( {{\theta _u},{\varphi _u}} \right){{\bf{\tilde{a}}}_t}\left( {{\bm{\theta }},{\bm{\varphi }}} \right){\bf{B}}\left( {{{\bm{\beta }}_u}} \right)\left( {{\bf{Vs}} + {\tilde{\bf{r}}}} \right) + {{\bf{n}}_u}.
\end{equation}
Then, each auxiliary LEO satellite sends its received signal ${{\bf{y}}_u}$ to the central satellite via inter-satellite links for data fusion and collaborative target location estimation\footnote{Note that inter-satellite links may have an impact on both information communication and location sensing functions. However, the LEO satellite constellations are deploying inter-satellite optical links, which are able to provide high-speed signal transmission. For example, the inter-satellite optical links in the Starlink system can achieve transmission rates exceeding 100 Gbps \cite{100Gbps}. This means that the inter-satellite link communication capacity can meet the information exchange requirements of the current scenario and does not limit the data exchange between the central and auxiliary satellites, thus perfectly supporting the multiple-satellite cooperative information communication and location sensing in LEO satellite constellations.}. The aggregated signal ${\bf{y}} = \left[ {{{\bf{y}}_1}; \cdots ;{{\bf{y}}_K}} \right] \in {\mathbb{C}^{NK \times 1}}$ at the central satellite can be expressed as
\begin{equation}
{\bf{y}} = \alpha {\bf{A}}\left( {{\bm{\theta }},{\bm{\varphi }}} \right) \odot {\rm{ }}{\bf{B}}\left( {\bm{\beta }} \right)\left( {{\bf{Vs}} + {\tilde{\bf{r}}}} \right) + {\bf{n}},
\end{equation}
where ${\bf{A}}\left( {{\bm{\theta }},{\bm{\varphi }}} \right)= \left[ {{{\bf{a}}_r}\left( {{\theta _1},{\varphi _1}} \right); \cdots ;{{\bf{a}}_r}\left( {{\theta _K},{\varphi _K}} \right)} \right]\cdot{{\bf{\tilde{a}}}_t}\left( {{\bm{\theta }},{\bm{\varphi }}} \right) \in {\mathbb{C}^{NK \times NK}}$, ${\bf{B}}\left( {\bm{\beta }} \right) = {\left[ {{{\bm{\beta }}_1}, \cdots ,{{\bm{\beta }}_K}} \right]^{\rm T}} \otimes {{\bf{1}}_{N \times N}} \in {\mathbb{C}^{NK \times NK}}$ with ${\bm{\beta }} = \left[ {{\beta _{1,1}}; \cdots ;{\beta _{1,K}}; \cdots ;{\beta _{K,K}}} \right]$ and ${{\bf{1}}_{N \times N}}$ being an $N$-dimensional square matrix with all elements are equal to 1, and ${\bf{n}} = \left[ {{{\bf{n}}_1}; \cdots ;{{\bf{n}}_K}} \right] \in {\mathbb{C}^{NK \times 1}}$ satisfies $\mathbf{n}\sim\mathcal{C}\mathcal{N}\left( {0,\sigma_n ^2\mathbf{I}_{NK}} \right)$ .

With ${\bf{y}}$ in hand, we employ a maximum likelihood estimation (MLE) scheme to estimate the location of the target. In particular, since the target coordinate vector $\mathbf{p}$ and the reflection coefficient $\alpha$ are both unknown parameters, the likelihood function of {\bf{y}} can be expressed as \cite{MLE ref}
\begin{align}
&f\left( {{\bf{y}};{\bf{p}},\alpha } \right) = \frac{1}{{\sqrt {{{\left( {\pi {\sigma_n ^2}} \right)}^{NK}}} }}\cdot\notag \\
&\ \ \ \ \ \ \ \exp \left( { - \frac{1}{{{\sigma ^2}}}{{\left\| {{\bf{y}} - \alpha {\tilde{\bf{A}}}\left( {\bf{p}} \right) \odot {\bf{B}}\left( {\bm{\beta }} \right)\left( {{\bf{Vs}} + {\tilde{\bf{r}}}} \right)} \right\|}^2}} \right),
\end{align}
where ${\tilde{\bf{A}}}\left( {\bf{p}} \right)$ is obtained from ${\bf{A}}\left( {{\bm{\theta }},{\bm{\varphi }}} \right)$ according to the mapping relation equations (\ref{theta}) and (\ref{varphi}) for angles and coordinates. Accordingly, the MLE of $\left(\mathbf{p},\alpha\right)$ is given by
\begin{align}\label{all_MLE}
&\left( {{{\bf{p}}_{{\rm{MLE}}}},{\alpha _{{\rm{MLE}}}}} \right) = \arg \mathop {\max }\limits_{{\bf{p}},\alpha } f\left( {{\bf{y}};{\bf{p}},\alpha } \right) \notag \\
&\ \ \ \ \ \ \ \ = \arg \mathop {\min }\limits_{{\bf{p}},\alpha } {\left\| {{\bf{y}} - \alpha {\tilde{\bf{A}}}\left( {\bf{p}} \right) \odot {\bf{B}}\left( {\bf{\beta }} \right)\left( {{\bf{Vs}} + {\tilde{\bf{r}}}} \right)} \right\|^2}.
\end{align}
For given {\bf{p}}, the MLE of $\alpha$ can be determined as
\begin{align}\label{alpha_MLE}
{\alpha _{{\rm{MLE}}}} &= \arg \mathop {\min }\limits_\alpha  {\left\| {{\bf{y}} - \alpha {\tilde{\bf{A}}}\left( {\bf{p}} \right) \odot {\bf{B}}\left( {\bf{\beta }} \right)\left( {{\bf{Vs}} + {\tilde{\bf{r}}}} \right)} \right\|^2}\notag \\
&= \frac{{{{\left( {{\bf{Vs}} + {\tilde{\bf{r}}}} \right)}^{\rm H}}\left( {{{\tilde{\bf{A}}}^{\rm H}}\left( {\bf{p}} \right) \odot {{\bf{B}}^{\rm H}}\left( {\bm{\beta }} \right)} \right){\bf{y}}}}{{{{\left\| {{\tilde{\bf{A}}}\left( {\bf{p}} \right) \odot {\bf{B}}\left( {\bm{\beta }} \right)\left( {{\bf{Vs}} + {\tilde{\bf{r}}}} \right)} \right\|}^2}}}.
\end{align}
Inserting ${\alpha _{{\rm{MLE}}}}$ into equation (\ref{all_MLE}), we obtain
\begin{align}\label{alpha_MLE2}
&{\left\| {{\bf{y}} - {\alpha _{{\rm{MLE}}}}{\tilde{\bf{A}}}\left( {\bf{p}} \right) \odot {\bf{B}}\left( {\bm{\beta }} \right)\left( {{\bf{Vs}} + {\tilde{\bf{r}}}} \right)} \right\|^2}  \notag \\
&= {\left\| {\bf{y}} \right\|^2} - \frac{{{{\left| {{{\left( {{\bf{Vs}} + {\tilde{\bf{r}}}} \right)}^{\rm H}}\left( {{{\tilde{\bf{A}}}^{\rm H}}\left( {\bf{p}} \right) \odot {{\bf{B}}^{\rm H}}\left( {\bm{\beta }} \right)} \right){\bf{y}}} \right|}^2}}}{{{{\left\| {{\tilde{\bf{A}}}\left( {\bf{p}} \right) \odot {\bf{B}}\left( {\bm{\beta }} \right)\left( {{\bf{Vs}} + {\tilde{\bf{r}}}} \right)} \right\|}^2}}}.
\end{align}
As a result, after substituting (\ref{alpha_MLE2}) into (\ref{all_MLE}), the MLE of {\bf{p}} is given by
\begin{equation}\label{p_MLE}
{{\bf{p}}_{{\rm{MLE}}}} = \arg \mathop {\max }\limits_{\bf{p}} \frac{{{{\left| {{{\left( {{\bf{Vs}} + {\tilde{\bf{r}}}} \right)}^{\rm H}}\left( {{{\tilde{\bf{A}}}^{\rm H}}\left( {\bf{p}} \right) \odot {{\bf{B}}^{\rm H}}\left( {\bm{\beta }} \right)} \right){\bf{y}}} \right|}^2}}}{{{{\left\| {{\tilde{\bf{A}}}\left( {\bf{p}} \right) \odot {\bf{B}}\left( {\bm{\beta }} \right)\left( {{\bf{Vs}} + {\tilde{\bf{r}}}} \right)} \right\|}^2}}},
\end{equation}
whose solution can usually be found by the grid search method. However, when tackling large-scale spatial location estimation, the computational complexity of the grid search method based on the exhaustive search principle shows an exponential growth, which cannot satisfy the requirements of high-precision real-time location sensing. To overcome this challenge, we adopt a low-complexity PSO-based direct location sensing algorithm to solve the non-convex location estimation problem (\ref{p_MLE}). Specifically, the PSO method is an evolutionary computation technique that originated from the study of bird swarm feeding behavior, aiming to find the optimal solution through collaboration and information sharing among individuals in the swarm. It simulates birds in a swarm by designing massless particles with only two attributes: velocity and location, where velocity represents the speed of movement and location represents the direction of movement.
Let $I_p$ denote the number of particles, $N_p$ denote the maximum number of iterations, $i_p$ denote particle index, $n_p$ denote iteration index, ${\bf{p}}_{{i_p}}^{\left( {{n_p}} \right)}\in \mathbb{R}^{3\times1}$ and ${\bf{v}}_{{i_p}}^{\left( {{n_p}} \right)}\in \mathbb{R}^{3\times1}$ denote the location and velocity of the $i_p$-th particle in three-dimensional (3D) space at the $n_p$-th iteration, respectively. Moreover, $\varpi$ stands for the  inertia weight using linearly decreasing weight (LDW) strategy, which offers a more balanced capability for both local and global optimization compared to a fixed weight.
In this algorithm, particles perform intelligent search through information exchange and cooperation. The MLE function of $\mathbf{p}$ is utilized as the fitness function to characterize the accuracy of location estimation. By continuously updating velocity ${\bf{v}}_{{i_p}}^{\left( {{n_p}} \right)}$ and location ${\bf{p}}_{{i_p}}^{\left( {{n_p}} \right)}$ of each particle with reference to the experience of historical optimal location of individual ${{\bf{p}}{{_{{i_p}}^{{\rm{best}}}}}}$ and swarm ${{{\bf{g}}^{{\rm{best}}}}}$, the direct location sensing algorithm gradually approaches the global optimum $\widehat{{\bf{p}}}$. The detailed procedures of the algorithm are listed in Algorithm 1.
\begin{algorithm} 
\caption{: PSO-based Direct Location Sensing Algorithm}
\label{alg1}
\hspace*{0.02in} {\bf Input:} 
 $\mathbf{y},{{\bf{q}}_k}, {{\bf{w}}_{k,m}}, {{\bf{r}}_k}, {\beta_{k,u}}, I_p, N_p $.\\
\hspace*{0.02in} {\bf Output:} 
$\widehat{{\bf{p}}}$.
\begin{algorithmic}[1]
\STATE{\textbf{Initialize} particle index $i_p=1$, iteration index $n_p=0$, particle location ${\bf{p}}_{{i_p}}^{\left( {{0}} \right)}, \forall i_p$, particle velocity ${\bf{v}}_{{i_p}}^{\left( {{0}} \right)}, \forall i_p$, individual learning factor $c_1$, swarm learning factor $c_2$, random numbers $r_{1},r_{2}\in[0,1]$, maximum inertia weight ${{\varpi _{\max }}}$, minimum inertia weight ${{\varpi _{\min}}}$,  fitness value $F({{\bf{p}}{{_{{i_p}}^{{\rm{best}}}}}})= F\left({{{\bf{g}}^{{\rm{best}}}}}\right)= -\infty,\forall i_p $.
}
\WHILE{$n_p \leq N_p$}
\FOR{$i_p=1:I_p$}
\STATE {Calculate fitness value of the $i_p$-th particle according to fitness function :\\
$F\left({\bf{p}}_{{i_p}}^{\left( {{n_p}} \right)}\right)=\frac{{{{\left| {{{\left( {{\bf{Vs}} + {\tilde{\bf{r}}}} \right)}^{\rm H}}\left( {{{\tilde{\bf{A}}}^{\rm H}}\left( {\bf{p}}_{{i_p}}^{\left( {{n_p}} \right)} \right) \odot {{\bf{B}}^{\rm H}}\left( {\bm{\beta }} \right)} \right){\bf{y}}} \right|}^2}}}{{{{\left\| {{\tilde{\bf{A}}}\left( {\bf{p}}_{{i_p}}^{\left( {{n_p}} \right)} \right) \odot {\bf{B}}\left( {\bm{\beta }} \right)\left( {{\bf{Vs}} + {\tilde{\bf{r}}}} \right)} \right\|}^2}}}$;}
\IF{$F\left({\bf{p}}_{{i_p}}^{\left( {{n_p}} \right)}\right)>F\left( {{\bf{p}}{{_{{i_p}}^{{\rm{best}}}}}}\right)$}
\STATE{Update historical optimal location of the $i_p$-th particle :
${{\bf{p}}{{_{{i_p}}^{{\rm{best}}}}}}={\bf{p}}_{{i_p}}^{\left( {{n_p}} \right)}$;}
\ENDIF
\IF{${\bf{p}}_{{i_p}}^{\left( {{n_p}} \right)}>{{{\bf{g}}^{{\rm{best}}}}}$}
\STATE{Update historical optimal location of the particle swarm: ${{{\bf{g}}^{{\rm{best}}}}}={\bf{p}}_{{i_p}}^{\left( {{n_p}} \right)}$;}
\ENDIF
\ENDFOR
\FOR{$i_p=1:I_p$}
\STATE {Update the velocity of the $i_p$-th particle:\\ ${\bf{v}}_{{i_p}}^{\left( {{n_p} + 1} \right)} = \varpi {\bf{v}}_{{i_p}}^{\left( {{n_p}} \right)} + {c_1}{r_1}\left( {{\bf{p}}{{_{{i_p}}^{{\rm{best}}}}} - {\bf{p}}_{{i_p}}^{\left( {{n_p}} \right)}} \right) + {c_2}{r_2}\left( {{{\bf{g}}^{{\rm{best}}}} - {\bf{p}}_{{i_p}}^{\left( {{n_p}} \right)}} \right)$};
\STATE {Update the location of the $i_p$-th particle:\\ ${\bf{p}}_{{i_p}}^{\left( {{n_p} + 1} \right)} = {\bf{p}}_{{i_p}}^{\left( {{n_p}} \right)} + {\bf{v}}_{{i_p}}^{\left( {{n_p}} \right)}$};
\ENDFOR
\STATE{Update inertia weight: $\varpi  = {\varpi _{\max }} - \frac{{\left( {{\varpi _{\max }} - {\varpi _{min}}} \right){n_p}}}{{{N_p}}}$;}
\STATE {Update iteration index: $n_p=n_p+1$};
\ENDWHILE
\STATE{Obtain estimated target location $\widehat{{\bf{p}}}={{{\bf{g}}^{{\rm{best}}}}}$.}
\end{algorithmic}
\end{algorithm}

Herein, we analyze the performance of the proposed PSO-based direct location sensing algorithm in terms of CRB. Specifically, CRB represents the theoretical limit for parameter estimation, indicating the minimum achievable variance. Considering the observed aggregated signal $\mathbf{y}$ and model information, the variance of any unbiased estimator cannot be lower than this bound. It is important to note that the CRB is solely dependent on the observed signal and model, and is independent of the specific localization approach employed. In general, the diagonal elements of the CRB matrix represent the minimum variance of the coordinate estimation with respect to the x, y, and z axes, respectively \cite{CRB ref}. Therefore, to evaluate the performance of the target location estimation in the 3D coordinate system, we focus on the trace of the CRB matrix ${\bf{C}}\in {\mathbb{R}^{3 \times 3}}$, which can be expressed as \cite{CRB ref2}
\begin{equation}\label{C}
{\rm{tr}}\left( {{\bf{C}}} \right) = {\rm{tr}}\left( {{\bf{F}}_{\bf{p}}^{-1}} \right){\rm{ = tr}}\left( {{{\left( {{\bf{J}}^{\rm T}{{\bf{F}}_{\bf{\Omega }}}{{\bf{J}}}} \right)}^{ - 1}}} \right),
\end{equation}
where ${{\bf{F}}_{\bf{p}}}\in {\mathbb{R}^{3 \times 3}}$ denotes the Fisher information matrix (FIM) of location vector $\bf{p}$, ${{{\bf{F}}_{\bm{\Omega }}}}\in {\mathbb{R}^{2K \times 2K}}$ stands for the FIM of angle vector ${\bf{\Omega }} = \left[ {{\bm{\theta }};{\bm{\varphi }}} \right] \in {\mathbb{R}^{2K \times 1}}$, and ${{\bf{J}}} \in {\mathbb{R}^{2K \times 3}}$ denotes the Jacobian matrix of the linear mapping relation from $\bf{\Omega}$ to $\bf{p}$. The detailed calculation of ${{\bf{J}}}$ is shown in Appendix A. To further derive the specific expression for the CRB from ${{\bf{F}}_{\bf{\Omega }}}$, we examine the FIM with respect to the vector ${\bf{\Xi }} = \left[ {{\bf{\Omega }};\alpha } \right]\in {\mathbb{R}^{(2K+1) \times 1}}$ with all unknown parameters, which is given by
\begin{equation}
{{\bf{F}}_{\bf{\Xi }}} = \left[ {\begin{array}{*{20}{c}}
{{{\bf{F}}_{{\bf{\Omega \Omega }}}}}&{{{\bf{F}}_{{\bf{\Omega }}\alpha }}}\\
{{{\bf{F}}_{{\bf{\Omega }}\alpha }^{\rm T}}}&{{{\bf{F}}_{\alpha \alpha }}}
\end{array}} \right],
\end{equation}
where
\begin{align}\label{F_xi}
&{\left[ {{{\bf{F}}_{\bf{\Xi }}}} \right]_{i,j}} \notag\\
&\ \ \ = 2{\mathop{\rm Re}\nolimits} \left\{ {\frac{{\partial {{\bf{u}}^{\rm H}}}}{{\partial {[{\bf{\Xi}}] _i}}}{\bf{R}}_n^{ - 1}\frac{{\partial {\bf{u}}}}{{\partial {[{\bf{\Xi}}] _j}}}} \right\} + {\rm{tr}}\left( {{\bf{R}}_n^{ - 1}\frac{{\partial {{\bf{R}}_n}}}{{\partial {[{\bf{\Xi}}] _i}}}{\bf{R}}_n^{ - 1}\frac{{\partial {{\bf{R}}_n}}}{{\partial {[{\bf{\Xi}}] _j}}}} \right) \notag\\
&\ \ \ = \frac{2}{{\sigma _n^2}}{\mathop{\rm Re}\nolimits} \left\{ {\frac{{\partial {{\bf{u}}^{\rm H}}}}{{\partial {[{\bf{\Xi}}] _i}}}\frac{{\partial {\bf{u}}}}{{\partial {[{\bf{\Xi}}] _j}}}} \right\}
\end{align}
with $\mathbf{u}=\alpha {\bf{A}}\left( {{\bm{\theta }},{\bm{\varphi }}} \right) \odot {\rm{ }}{\bf{B}}\left( {\bm{\beta }} \right)\left( {{\bf{Vs}} + {\tilde{\bf{r}}}} \right)$ and ${\bf{R}}_n=\sigma_n ^2\mathbf{I}_{NK}$ \cite{FIM}.
In this context, we have \cite{FIM2}
\begin{equation}\label{F_O}
{{\bf{F}}_{\bf{\Omega }}} = {{\bf{F}}_{{\bf{\Omega \Omega }}}} - {{\bf{F}}_{{\bf{\Omega }}\alpha }}{\bf{F}}_{\alpha \alpha }^{ - 1}{\bf{F}}_{{\bf{\Omega }}\alpha }^{\rm T}.
\end{equation}
Next, we derive detailed expressions for ${{\bf{F}}_{{\bf{\Omega \Omega }}}}$, ${\bf{F}}_{\alpha \alpha }$ and ${\bf{F}}_{{\bf{\Omega }}\alpha }$.
By simplifying ${\bf{A}}\triangleq{\bf{A}}\left( {{\bm{\theta }},{\bm{\varphi }}} \right)$ and ${\bf{B}}\triangleq{\bf{B}}\left( {\bm{\beta }} \right)$, the partial derivatives of $\mathbf{u}$ with respect to unknown parameters are calculated as
\begin{equation}\label{u_Omega}
\frac{{\partial {\bf{u}}}}{{\partial {[{\bf{\Omega }}]_i}}} = \alpha \left( {\frac{{\partial {\bf{A}}}}{{\partial {[{\bf{\Omega }}]_i}}} \odot {\bf{B}}} \right)\left( {{\bf{Vs}} + {\tilde{\bf{r}}}} \right)\in {\mathbb{C}^{NK \times 1}}
\end{equation}
and
\begin{equation}
\frac{{\partial {\bf{u}}}}{{\partial \alpha }} = {\bf{A}} \odot {\bf{B}}\left( {{\bf{Vs}} + {\tilde{\bf{r}}}} \right)\in {\mathbb{C}^{NK \times 1}}.
\end{equation}
In particular, there is a compact form for equation (\ref{u_Omega}), namely
\begin{align}
&\frac{{\partial {\bf{u}}}}{{\partial {\bf{\Omega }}}} = \alpha \bigg( \frac{{\partial {\bf{A}}}}{{\partial {{\theta} _1}}} \odot {\bf{B}}, \cdots ,\frac{{\partial {\bf{A}}}}{{\partial {{\theta} _K}}} \odot {\bf{B}},\frac{{\partial {\bf{A}}}}{{\partial {{\varphi} _1}}} \odot {\bf{B}},\cdots, \notag\\
& \ \ \ \frac{{\partial {\bf{A}}}}{{\partial {{\varphi} _K}}} \odot {\bf{B}} \bigg)\left( {{{\bf{I}}_{2K}} \otimes \left( {{\bf{Vs}} + {\tilde{\bf{r}}}} \right)} \right) \in {\mathbb{C}^{NK \times 2K}}.
\end{align}
Consequently, based on equation (\ref{F_xi}), we obtain
\begin{equation}\label{F_oo}
{{\bf{F}}_{{\bf{\Omega \Omega }}}} = \frac{2}{{\sigma _n^2}}{\mathop{\rm Re}\nolimits} \left\{ {{{\frac{{\partial {\bf{u}}}}{{\partial {\bf{\Omega }}}}}^{\rm H}}\frac{{\partial {\bf{u}}}}{{\partial {\bf{\Omega }}}}} \right\} = \frac{2}{{\sigma _n^2}}{\mathop{\rm Re}\nolimits} \left\{ {{{\bf{Y}}^{\rm H}}{{\bf{X}}^{\rm H}}{\bf{XY}}} \right\},
\end{equation}
\begin{equation}\label{F_aa}
{{\bf{F}}_{\alpha \alpha }} = \frac{2}{{\sigma _n^2}}{\mathop{\rm Re}\nolimits} \left\{ {{{\frac{{\partial {\bf{u}}}}{{\partial \alpha }}}^{\rm H}}\frac{{\partial {\bf{u}}}}{{\partial \alpha }}} \right\} = \frac{2}{{\sigma _n^2}}{\mathop{\rm Re}\nolimits} \left\{ {{{{\bf{\tilde Y}}}^{\rm H}}{{{\bf{\tilde X}}}^{\rm H}}{\bf{\tilde X\tilde Y}}} \right\},
\end{equation}
\begin{equation}\label{F_oa}
{{\bf{F}}_{{\bf{\Omega }}\alpha }} = \frac{2}{{\sigma _n^2}}{\mathop{\rm Re}\nolimits} \left\{ {{{\frac{{\partial {\bf{u}}}}{{\partial {\bf{\Omega }}}}}^{\rm H}}\frac{{\partial {\bf{u}}}}{{\partial \alpha }}} \right\} = \frac{2}{{\sigma _n^2}}{\mathop{\rm Re}\nolimits} \left\{ {{{\bf{Y}}^{\rm H}}{{\bf{X}}^{\rm H}}{\bf{\tilde X\tilde Y}}} \right\},
\end{equation}
where ${\bf{Y}}=\left( {{{\bf{I}}_{2K}} \otimes \left( {{\bf{Vs}} + {\tilde{\bf{r}}}} \right)} \right)$, ${\bf{\tilde Y}}=\left( {{\bf{Vs}} + {\tilde{\bf{r}}}} \right)$, $\mathbf{X}=\alpha \left( {\frac{{\partial {\bf{A}}}}{{\partial {{\theta }_1}}} \odot {\bf{B}}, \cdots ,\frac{{\partial {\bf{A}}}}{{\partial {{\theta} _K}}} \odot {\bf{B}},\frac{{\partial {\bf{A}}}}{{\partial {{\varphi} _1}}} \odot {\bf{B}}, \cdots ,\frac{{\partial {\bf{A}}}}{{\partial {{\varphi} _K}}} \odot {\bf{B}}} \right)$ and ${\bf{\tilde X}}={\bf{A}} \odot {\bf{B}}$.
Furthermore, since ${\rm E}[{\bf{s}}{{\bf{s}}^{\rm H}}] = {{\bf{I}}_M}$, ${\rm E}[{\bf{s}}{{\tilde{\bf{r}}}^{\rm H}}] = {{\bf{0}}_{M \times NK}}$ and $\left( {{{\bf{I}}_{2K}} \otimes \left( {{\bf{Vs}} + {\tilde{\bf{r}}}} \right)} \right) = \left( {{{\bf{I}}_{2K}} \otimes \left( {\sum\limits_{m = 1}^M {{\tilde{{\bf{w}}}_m}{s_m}} } \right)} \right) + \left( {{{\bf{I}}_{2K}} \otimes {\tilde{\bf{r}}}} \right)$, the elements of the matrices in equations (\ref{F_oo})-(\ref{F_oa}) can be written as
\begin{align}\label{F_oo_ele}
&{\left[ {{{\bf{F}}_{{\bf{\Omega \Omega }}}}} \right]_{i,j}} = \frac{{2{\alpha ^2}}}{{\sigma _n^2}}{\mathop{\rm Re}\nolimits} \Bigg( {{\tilde{\bf{r}}}^{\rm H}}\left( {{{\frac{{\partial {\bf{A}}}}{{\partial {[{\bf{\Omega }}]_i}}}}^{\rm H}} \odot {{\bf{B}}^{\rm H}}} \right)\left( {\frac{{\partial {\bf{A}}}}{{\partial {[{\bf{\Omega }}]_j}}} \odot {\bf{B}}} \right){\tilde{\bf{r}}} \notag\\
&\ \ \ \ \ \ + \sum\limits_{m = 1}^M {\tilde{{\bf{w}}}_m^{\rm H}\left( {{{\frac{{\partial {\bf{A}}}}{{\partial {[{\bf{\Omega }}]_i}}}}^{\rm H}} \odot {{\bf{B}}^{\rm H}}} \right)\left( {\frac{{\partial {\bf{A}}}}{{\partial {[{\bf{\Omega }}]_j}}} \odot {\bf{B}}} \right){\tilde{{\bf{w}}}_m}}  \Bigg),
\end{align}
\begin{align}\label{F_aa_ele}
&{{\bf{F}}_{\alpha \alpha }} = \frac{2}{{\sigma _n^2}}{\mathop{\rm Re}\nolimits} \Bigg( {{\tilde{\bf{r}}}^{\rm H}}\left( {{{\bf{A}}^{\rm H}} \odot {{\bf{B}}^{\rm H}}} \right)\left( {{\bf{A}} \odot {\bf{B}}} \right){\tilde{\bf{r}}}\notag\\
&\ \ \ \ \ \ \ \ \ \ \ \ \ \ + \sum\limits_{m = 1}^M {\tilde{{\bf{w}}}_m^{\rm H}\left( {{{\bf{A}}^{\rm H}} \odot {{\bf{B}}^{\rm H}}} \right)\left( {{\bf{A}} \odot {\bf{B}}} \right){\tilde{{\bf{w}}}_m}}  \Bigg)
\end{align}
and
\begin{align}\label{F_oa_ele}
&{\left[ {{{\bf{F}}_{{\bf{\Omega }}\alpha }}} \right]_i} = \frac{{2\alpha }}{{\sigma _n^2}}{\mathop{\rm Re}\nolimits} \Bigg( {{\tilde{\bf{r}}}^{\rm H}}\left( {{{\frac{{\partial {\bf{A}}}}{{\partial {[{\bf{\Omega }}]_i}}}}^{\rm H}} \odot {{\bf{B}}^{\rm H}}} \right)\left( {{\bf{A}} \odot {\bf{B}}} \right){\tilde{\bf{r}}} \notag\\
&\ \ \ \ \ \ \ \ \ + \sum\limits_{m = 1}^M {\tilde{{\bf{w}}}_m^{\rm H}\left( {{{\frac{{\partial {\bf{A}}}}{{\partial {[{\bf{\Omega }}]_i}}}}^{\rm H}} \odot {{\bf{B}}^{\rm H}}} \right)\left( {{\bf{A}} \odot {\bf{B}}} \right){\tilde{{\bf{w}}}_m}}  \Bigg).
\end{align}

From equations (\ref{C}), (\ref{F_O}) and (\ref{F_oo_ele})-(\ref{F_oa_ele}), it is observed that the CRB as a performance metric for location sensing has an explicit closed-form expression and its values are related to the cooperative communication beamforming ${\tilde{{\bf{w}}}_{m}}$ and the cooperative sensing waveform ${{\tilde{\bf{r}}}}$. Therefore, it is feasible to improve the cooperative location sensing performance by designing suitable ${\tilde{{\bf{w}}}_{m}}$ and ${{\tilde{\bf{r}}}}$.

\section{Design of Multiple-satellite Cooperative Information Communication and Location Sensing}
In earlier performance analysis, it is found that the communication beamforming and sensing waveform have great impacts on the system performance. In this section, we propose a multiple-satellite cooperative information communication and location sensing design for dual-function LEO satellite constellation by jointly optimizing communication beamforming and sensing waveform.
\subsection{Problem Formulation}
To enhance the overall performance of dual-function LEO satellite constellation, we aim to minimize the trace of the CRB matrix while guaranteeing the quality of information communication and meeting transmit power constraints. In particular, the design can be formulated as the following optimization problem:
\begin{subequations}\label{OP1}
\begin{align}
\mathop {\min }\limits_{{\tilde{{\bf{w}}}_m},{\tilde{\bf{r}}}}&\ \ {\rm{ tr}}\left( {{{\left( {{{\bf{J}}^{\rm T}}\left({{{\bf{F}}_{{\bf{\Omega \Omega }}}} - {{\bf{F}}_{{\bf{\Omega }}\alpha }}{\bf{F}}_{\alpha \alpha }^{ - 1}{\bf{F}}_{{\bf{\Omega }}\alpha }^{\rm T}}\right){\bf{J}}} \right)}^{ - 1}}} \right)
\label{OP1obj} \\
{\rm{s.t.}}
&\ \ \sum\limits_{m = 1}^M {{{\left\| {{{\bf{\Lambda }}_k}{\tilde{{\bf{w}}}_{m}}} \right\|}^2}}  + {{{\left\| {{{\bf{\Lambda }}_k}{{\tilde{\bf{r}}}}} \right\|}^2}} \le P_k^{\max }, \label{OP1st1} \\
&\ \ {R_i} \ge {\eta _i}, \label{OP1st2}
\end{align}
\end{subequations}
where ${{\bf{\Lambda }}_k} = {\bm{\gamma }}_k \otimes {{\bf{I}}_N}\in {\mathbb{R}^{N \times NK}}$ with ${\bm{\gamma }}_k$ being a $K$-dimensional row vector whose $k$-th element is 1, and the remaining elements are 0. Furthermore, the objective function (\ref{OP1obj}) is derived from equations (\ref{C}) and (\ref{F_O}), aiming to improve the location sensing accuracy, constraint (\ref{OP1st1}) based on equation (\ref{power}) signifies the transmit power limitation of LEO satellites with $P_k^{\max }$ being the maximum transmit power budget of the $k$-th LEO satellite, and constraint (\ref{OP1st2}) puts requirements on the quality of information communication service for the UEs served by the LEO satellite constellation with ${\eta _i}$ being the required transmission rate of the $i$-th UE. It should be noted that problem (\ref{OP1}) involves quadratic terms of the optimization variables and complex matrix operations, making it impossible to solve for the optimal solution within polynomial time \cite{QCQP}. For this reason, we explore an efficient algorithm to obtain feasible suboptimal solutions, ultimately enhancing the overall performance of multiple-satellite cooperative information communication and location sensing for LEO satellite constellations.

\subsection{Algorithm Design}
To tackle the non-convexity of objective function (\ref{OP1obj}), we introduce an auxiliary variable matrix ${\bf{U}}\in {\mathbb{R}^{3 \times 3}}$ with ${ {{{\bf{J}}^{\rm T}}\left({{{\bf{F}}_{{\bf{\Omega \Omega }}}} - {{\bf{F}}_{{\bf{\Omega }}\alpha }}{\bf{F}}_{\alpha \alpha }^{ - 1}{\bf{F}}_{{\bf{\Omega }}\alpha }^{\rm T}}\right){\bf{J}}} } \succeq{\bf{U}}$. Specifically, according to the Schur complement theorem \cite{Schur}, the objective function can be equivalent to minimizing ${\rm{tr}}\left( {{{\bf{U}}^{ - 1}}} \right)$, which subjects to the constraint
\begin{equation}\label{U}
\left[ {\begin{array}{*{20}{c}}
{{{\bf{J}}^{\rm T}}{{\bf{F}}_{{\bf{\Omega \Omega }}}}{\bf{J}} - {\bf{U}}}&{{{\bf{J}}^{\rm T}}{{\bf{F}}_{{\bf{\Omega }}\alpha }}}\\
{{\bf{F}}_{{\bf{\Omega }}\alpha }^{\rm T}{\bf{J}}}&{{\bf{F}}_{\alpha \alpha }^{}}
\end{array}} \right] \succeq 0.
\end{equation}
Nevertheless, inequality (\ref{U}) remains non-convex due to the presence of quadratic terms in the elements of ${{\bf{F}}_{{\bf{\Omega \Omega }}}}$, ${\bf{F}}_{\alpha \alpha }$ and ${\bf{F}}_{{\bf{\Omega }}\alpha }$. To address this issue, we apply the semi-definite relaxation (SDR) technique and define ${{\bf{W}}_m} = {\tilde{{\bf{w}}}_m}\tilde{{\bf{w}}}_m^{\rm H}$ and ${\bf{R}} = {\tilde{\bf{r}}}{{\tilde{\bf{r}}}^{\rm H}}$, such that ${\left[ {{{\bf{F}}_{{\bf{\Omega \Omega }}}}} \right]_{i,j}}$, ${{\bf{F}}_{\alpha \alpha }}$ and ${\left[ {{{\bf{F}}_{{\bf{\Omega }}\alpha }}} \right]_i}$ can be rephrased as
\begin{align}
&{\left[ {{{\bf{F}}_{{\bf{\Omega \Omega }}}}} \right]_{i,j}} = \frac{{2{\alpha ^2}}}{{\sigma _n^2}}{\rm{Tr}}\Bigg( {\bf{R}}\left( {{{\frac{{\partial {\bf{A}}}}{{\partial {[{\bf{\Omega }}]_i}}}}^{\rm H}} \odot {{\bf{B}}^{\rm H}}} \right)\left( {\frac{{\partial {\bf{A}}}}{{\partial {[{\bf{\Omega }}]_j}}} \odot {\bf{B}}} \right)\notag\\
&\ \ \ \ \ + \sum\limits_{m = 1}^M {{{\bf{W}}_m}\left( {{{\frac{{\partial {\bf{A}}}}{{\partial {[{\bf{\Omega }}]_i}}}}^{\rm H}} \odot {{\bf{B}}^{\rm H}}} \right)\left( {\frac{{\partial {\bf{A}}}}{{\partial {[{\bf{\Omega }}]_j}}} \odot {\bf{B}}} \right)} \Bigg),
\end{align}
\begin{align}
&{{\bf{F}}_{\alpha \alpha }} = \frac{2}{{\sigma _n^2}}{\rm{Tr}}\Bigg( {\bf{R}}\left( {{{\bf{A}}^{\rm H}} \odot {{\bf{B}}^{\rm H}}} \right)\left( {{\bf{A}} \odot {\bf{B}}} \right) \notag\\
&\ \ \ \ \ \ \ \ \ \ \ \ \ \ + \sum\limits_{m = 1}^M {{{\bf{W}}_m}\left( {{{\bf{A}}^{\rm H}} \odot {{\bf{B}}^{\rm H}}} \right)\left( {{\bf{A}} \odot {\bf{B}}} \right)}  \Bigg)
\end{align}
and
\begin{align}
&{\left[ {{{\bf{F}}_{{\bf{\Omega }}\alpha }}} \right]_i} = \frac{{2\alpha }}{{\sigma _n^2}}{\rm{Tr}}\Bigg( {\bf{R}}\left( {{{\frac{{\partial {\bf{A}}}}{{\partial {[{\bf{\Omega }}]_i}}}}^{\rm H}} \odot {{\bf{B}}^{\rm H}}} \right)\left( {{\bf{A}} \odot {\bf{B}}} \right) \notag\\
&\ \ \ \ \ \ \ + \sum\limits_{m = 1}^M {{{\bf{W}}_m}\left( {{{\frac{{\partial {\bf{A}}}}{{\partial {[{\bf{\Omega }}]_i}}}}^{\rm H}} \odot {{\bf{B}}^{\rm H}}} \right)\left( {{\bf{A}} \odot {\bf{B}}} \right)}  \Bigg).
\end{align}
Thus, constraint (\ref{U}) can be regarded as a standard linear matrix inequality (LMI).
Accordingly, the transmit power constraint (\ref{OP1st1}) is rewritten as
\begin{equation}\label{Pmax}
{\rm{tr}}\left( { \sum\limits_{m = 1}^M {{{\bf{\Lambda }}_k}{{\bf{W}}_m}{\bf{\Lambda }}_k^{\rm T}}+{{\bf{\Lambda }}_k}{\bf{R\Lambda }}_k^{\rm T} } \right) \le P_k^{\max }.
\end{equation}
Moreover, for transmission rate constraint (\ref{OP1st2}), let ${{\bf{H}}_i} = {\tilde{{\bf{h}}}_i}\tilde{{\bf{h}}}_i^{\rm H}$, then we have the following equivalent form
\begin{equation}\label{OP1st2_1}
{R_i} = {\log _2}\left( {1 + \frac{{{\rm{tr}}\left( {{{\bf{W}}_i}{{\bf{H}}_i}} \right)}}{{\sum\limits_{m \ne i}^M {{\rm{tr}}\left( {{{\bf{W}}_m}{{\bf{H}}_i}} \right)}  + {\rm{tr}}\left( {{\bf{R}}{{\bf{H}}_i}} \right) + \sigma _i^2}}} \right) \ge {\eta _i}.
\end{equation}
To tackle the non-convex fractional constraint (\ref{OP1st2_1}), let ${Z'_i} =  - {\log _2}\left( {{\rm{tr}}\left( {{{\bf{W}}_i}{{\bf{H}}_i}} \right) + \sum\limits_{m \ne i}^M {{\rm{tr}}\left( {{{\bf{W}}_m}{{\bf{H}}_i}} \right)}  + {\rm{tr}}\left( {{\bf{R}}{{\bf{H}}_i}} \right) + \sigma _i^2} \right)$ and ${{\rm Z}''_i} =  - {\log _2}\left( {\sum\limits_{m \ne i}^M {{\rm{tr}}\left( {{{\bf{W}}_m}{{\bf{H}}_i}} \right)}  + {\rm{tr}}\left( {{\bf{R}}{{\bf{H}}_i}} \right) + \sigma _i^2} \right)$. By applying difference of convex algorithm (DCA) approach, we can obtain $ - {R_i} = {Z'_i} - {{\rm Z}''_i}$, where ${Z'_i}$ and ${{\rm Z}''_i}$ are both jointly convex functions with respect to the optimization variables ${{\bf{W}}_m}$ and ${\bf{R}}$. Further, we utilize the successive convex approximation (SCA) method on ${{\rm Z}''_i}$, whose first-order Taylor expansion at the $\left( {{\bf{W}}_m^\# ,{{\bf{R}}^\# }} \right)$ point can be written as
\begin{align}
&{\widetilde {\rm Z}''_i}\left( {{{\bf{W}}_m},{\bf{R}}} \right) = {{\rm Z}''_i}\left( {{\bf{W}}_m^\# ,{{\bf{R}}^\# }} \right)\notag\\
&\ \ \ \ \ \ \ \ + \sum\limits_{m \ne i}^M {\left\langle {\nabla _{{{\bf{W}}_m}}^{}\left( {{{\rm Z}''_i}\left( {{\bf{W}}_m^\# ,{{\bf{R}}^\# }} \right)} \right) \cdot \left( {{{\bf{W}}_m} - {\bf{W}}_m^\# } \right)} \right\rangle } \notag \\
&\ \ \ \ \ \ \ \ + \left\langle {\nabla _{\bf{R}}^{}\left( {{{{\rm Z}_i''}}\left( {{\bf{W}}_m^\# ,{{\bf{R}}^\# }} \right)} \right) \cdot \left( {{\bf{R}} - {{\bf{R}}^\# }} \right)} \right\rangle,\label{Z Taylor}
\end{align}
where $\nabla _{{{\bf{W}}_m}}^{}\left( {{{\rm Z}''_i}\left( {{\bf{W}}_m^\# ,{{\bf{R}}^\# }} \right)} \right) = \nabla _{\bf{R}}^{}\left( {{{\rm Z}''_i}\left( {{\bf{W}}_m^\# ,{{\bf{R}}^\# }} \right)} \right) =  - \frac{1}{{\ln 2}}\frac{{{\bf{H}}_i^{\rm T}}}{{\sum\limits_{m \ne i}^M {{\rm{tr}}\left( {{\bf{W}}_m^\# {{\bf{H}}_i}} \right)}  + {\rm{tr}}\left( {{{\bf{R}}^\# }{{\bf{H}}_i}} \right) + \sigma _i^2}}$, and $\left\langle  \cdot  \right\rangle $ represents the matrix inner product operation. Eventually, constraint (\ref{OP1st2_1}) is transformed as
\begin{equation}\label{OP1st2_2}
{\eta _i} + {Z'_i}\left( {{{\bf{W}}_m},{\bf{R}}} \right) - {\widetilde {\rm Z}''_i}\left( {{{\bf{W}}_m},{\bf{R}}} \right) \le 0.
\end{equation}
In particular, by exploiting the convexity of the function $-\log_2(\cdot)$ and the fact that affine mapping preserves convexity, it can be shown that ${Z'_i}\left( {{{\bf{W}}_m},{\bf{R}}} \right)$ is a convex function \cite{preserves convexity}. Moreover, based on equation (\ref{Z Taylor}), it can be inferred that ${\widetilde {\rm Z}''_i}\left( {{{\bf{W}}_m},{\bf{R}}} \right)$ is an affine function with respect to the variables $\left( {{{\bf{W}}_m},{\bf{R}}} \right)$. Consequently, it is evident that constraint (\ref{OP1st2_2}) represents a convex constraint. As a result, the original optimization problem (\ref{OP1}) is reformulated as
\begin{subequations}\label{OP2}
\begin{align}
\mathop {\min }\limits_{{{\bf{W}}_m},{\bf{R}},{\bf{U}}}&\ \ {\rm{ tr}}\left( {{{\bf{U}}^{ - 1}}} \right)\label{OP2obj} \\
{\rm{s.t.}}& \ \ (\ref{U}),(\ref{Pmax}),(\ref{OP1st2_2}), \notag\\
&\ \ \ {{\bf{W}}_m}\succeq0,  {\bf{R}}\succeq0, \label{OP2st1}\\
&\ \ \ {\rm{Rank}}\left( {{{\bf{W}}_m}} \right) = 1,  {\rm{Rank}}\left( {\bf{R}} \right) = 1, \label{OP2st2}
\end{align}
\end{subequations}
where (\ref{OP2st1}) and (\ref{OP2st2}) appear due to the utilization of SDR technique. Unfortunately, the optimization problem (\ref{OP2}) remains non-convex owing to the existence of the rank-one constraint (\ref{OP2st2}). To address this issue, we incorporate a penalty function into the objective function (\ref{OP2obj}) to enforce the rank-one condition, thereby allowing us to eliminate the constraint (\ref{OP2st2}). Specifically, as stated in equation (\ref{OP2st1}), both ${{\bf{W}}_m}$ and {\bf{R}} are positive semi-definite matrices with non-negative eigenvalues. Consequently, we have ${\rm{tr}}\left( {{{\bf{W}}_m}} \right) = \sum\limits_{i = 1}^{NK} {{\lambda _i}\left( {{{\bf{W}}_m}} \right)}  \ge {\lambda _{\max }}\left( {{{\bf{W}}_m}} \right)$, where ${\lambda _{i }}\left(\cdot \right)$ and ${\lambda _{\max }}\left(\cdot \right)$ denote the $i$-th eigenvalue and the maximum eigenvalue of a matrix, respectively. In this case, constraint ${\rm{Rank}}\left( {{{\bf{W}}_m}} \right) = 1$ and constraint ${\rm{tr}}\left( {{{\bf{W}}_m}} \right) - {\lambda _{\max }}\left( {{{\bf{W}}_m}} \right) = 0$ are equivalent and lead to the same conclusion regarding the variable $\bf{R}$. Inspired by this, by substituting the equational constraints with exterior penalty functions, the objective function of problem (\ref{OP2}) is updated as
\begin{align}\label{penalty 1}
&\mathop {\min }\limits_{{{\bf{W}}_m},{\bf{R}},{\bf{U}}} {\rm{ tr}}\left( {{{\bf{U}}^{ - 1}}} \right) + \rho \Bigg( \sum\limits_{m = 1}^M \Big( {{\rm{tr}}\left( {{{\bf{W}}_m}} \right) - {\lambda _{\max }}\left( {{{\bf{W}}_m}} \right)} \Big)\notag\\
&\ \ \ \ \ \ \ \ \ \ \ \ \ \ \ \ \ \ \ \  \ \ \ \ \ \ \ \ \ + \Big( {{\rm{tr}}\left( {\bf{R}} \right) - {\lambda _{\max }}\left( {\bf{R}} \right)} \Big)  \Bigg),
\end{align}
where $\rho$ denotes the penalty factor. Nevertheless, the objective function (\ref{penalty 1}) becomes non-convex due to the introduction of the non-smooth functions ${\lambda _{\max }}\left( {{{\bf{W}}_m}} \right)$ and ${\lambda _{\max }}\left( {\bf{R}} \right)$. To tackle this challenge, we employ an iterative approximation of the maximum eigenvalue to convert (\ref{penalty 1}) into a convex formulation. More specifically, we use the inequalities
\begin{align}\label{penalty 2}
&{\rm{tr}}\left( {{\bf{W}}_m^{\left( {j + 1} \right)}} \right) - {\left( {{\bf{v}}_{W,m}^{\left( j \right)}} \right)^{\rm H}}{\bf{W}}_m^{\left( {j + 1} \right)}{\bf{v}}_{W,m}^{\left( j \right)} \ge \notag\\
&\ \ \ \ \ \ \ \ \ {\rm{tr}}\left( {{\bf{W}}_m^{\left( {j + 1} \right)}} \right) - {\lambda _{\max }}\left( {{\bf{W}}_m^{\left( {j + 1} \right)}} \right) \ge 0
\end{align}
and
\begin{align}\label{penalty 3}
&{\rm{tr}}\left( {{\bf{R}}_{}^{\left( {j + 1} \right)}} \right) - {\left( {{\bf{v}}_R^{\left( j \right)}} \right)^{\rm H}}{\bf{R}}_m^{\left( {j + 1} \right)}{\bf{v}}_R^{\left( j \right)} \ge  \notag\\
&\ \ \ \ \ \ \ \ \ {\rm{tr}}\left( {{\bf{R}}_{}^{\left( {j + 1} \right)}} \right) - {\lambda _{\max }}\left( {{\bf{R}}_{}^{\left( {j + 1} \right)}} \right) \ge 0,
\end{align}
where ${{\bf{W}}_m^{\left( {j} \right)}}$ and ${{\bf{R}}^{\left( {j} \right)}}$ denote the values of ${\bf{W}}_m$ and ${\bf{R}}$ in the $j$-th iteration, respectively. Additionally,
${{\bf{v}}_{W,m}^{\left( j \right)}}$ and ${\bf{v}}_R^{\left( j \right)}$ represent the unit norm eigenvectors corresponding to the maximum eigenvalues ${\lambda _{\max }}\big( {{\bf{W}}_m^{\left( {j} \right)}} \big)$ and ${\lambda _{\max }}\left( {{\bf{R}}_{}^{\left( {j} \right)}} \right)$, respectively. As a result, the improved optimization problem at the $(j+1)$-th iteration is mathematically represented as (\ref{OP3obj}), shown at the top of the next page.
\begin{figure*}[!t]
\begin{align}
\mathop {\min }\limits_{{{\bf{W}}_m},{\bf{R}},{\bf{U}}}&\ \ {\rm{ tr}}\left( {{{\bf{U}}^{ - 1}}} \right) + \rho \left( {\sum\limits_{m = 1}^M {\left( {{\rm{tr}}\left( {{\bf{W}}_m^{\left( {j + 1} \right)}} \right) - {{\left( {{\bf{v}}_{W,m}^{\left( j \right)}} \right)}^{\rm H}}{\bf{W}}_m^{\left( {j + 1} \right)}{\bf{v}}_{W,m}^{\left( j \right)}} \right) + \left( {{\rm{tr}}\left( {{\bf{R}}_{}^{\left( {j + 1} \right)}} \right) - {{\left( {{\bf{v}}_R^{\left( j \right)}} \right)}^{\rm H}}{\bf{R}}^{\left( {j + 1} \right)}{\bf{v}}_R^{\left( j \right)}} \right)} } \right)\label{OP3obj} \\
{\rm{s.t.}}& \ \ (\ref{U}),(\ref{Pmax}),(\ref{OP1st2_2}),(\ref{OP2st1}).\notag
\end{align}
\hrulefill
\end{figure*}
Since optimization problem (\ref{OP3obj}) with the penalty function is a standard convex problem, it can be solved by off-the-shelf optimization toolkits \cite{CVX}. It is worth noting that the selection of the penalty factor $\rho$ is crucial. If $\rho$ is too large, it will increase the computing difficulties in minimizing the objective function. Conversely, if $\rho$ is too small, the minimum point of the objective function will deviate far from the optimal solution, leading to poor computing efficiency. Therefore, for the penalty factor, we employ a strictly increasing sequence during the iterations, defined as $\rho^{(j+1)}=\iota\rho^{(j)}$, where $\iota$ represents the amplification coefficient. Eventually, with the optimal solutions ${\bf{W}}_m^{*}$ and ${\bf{R}}^{*}$ obtained from iteration (\ref{OP3obj}), we can calculate the solution to the original problem $(\ref{OP1})$ by using the eigenvalue decomposition (EVD) method, i.e.,
\begin{equation}\label{EVD}
\tilde{{\bf{w}}}_m^* = \sqrt {{\lambda _{\max }}\left( {{\bf{W}}_m^*} \right)} {\bf{v}}_{W,m}^*,\  {{\tilde{\bf{r}}}^*} = \sqrt {{\lambda _{\max }}\left( {{{\bf{R}}^*}} \right)} {\bf{v}}_R^*,
\end{equation}
where $\mathbf{v}^{*}_{W,m}$ and $\mathbf{v}^{*}_{R}$ denote the unit norm eigenvectors corresponding to the ${{\lambda _{\max }}\left( {{\bf{W}}_m^*} \right)}$ and ${{\lambda _{\max }}\left( {{{\bf{R}}^*}} \right)}$, respectively.
Above all, the proposed penalty function-based design of multiple-satellite cooperative information communication and location sensing for LEO satellite constellations is summarized as Algorithm 2.

\begin{algorithm} 
\caption{: Design of Multiple-satellite Cooperative Information Communication and Location Sensing}
\label{alg1}
\hspace*{0.02in} {\bf Input:} 
 $K, M, N_x, N_z, \rho, \iota, {\eta _i}, P_k^{\max}$, $\Delta$.\\
\hspace*{0.02in} {\bf Output:} 
${\tilde{\bf{w}}}_m$, ${\tilde{\bf{r}}}$.
\begin{algorithmic}[1]
\STATE{\textbf{Initialize} iteration index $j=0$, initial feasible points ${{\bf{W}}_m^{\left( {0} \right)}}, \forall m$ and ${{\bf{R}}^{\left( {0} \right)}}$.}
\REPEAT
\STATE{Update ${\bf{W}}_m^\# = {{\bf{W}}_m^{\left( {j} \right)}}, {{\bf{R}}^\# } = {{\bf{R}}^{\left( {j} \right)}}$;}
\STATE{Obtain ${{\bf{W}}_m^{\left( {j+1} \right)}}$ and ${{\bf{R}}^{\left( {j+1} \right)}}$ by solving problem (\ref{OP3obj});}
\IF{${{\bf{W}}_m^{\left( {j+1} \right)}}$ and ${{\bf{R}}^{\left( {j+1} \right)}}$ converge}
\IF{$ \sum\limits_{m = 1}^M \left| {\rm{tr}}\left( {{\bf{W}}_m^{\left( {j + 1} \right)}} \right) - {\lambda _{\max }}\left({{\bf{W}}_m^{\left( {j + 1} \right)}}\right)\right|
+ \Big| {{\rm{tr}}\left( {{\bf{R}}_{}^{\left( {j + 1} \right)}} \right) - {\lambda _{\max }} \left({\bf{R}}^{\left( {j + 1} \right)}\right) } \Big|>\Delta
$}
\STATE{Update penalty factor $\rho=\iota\rho$};
\ENDIF
\ENDIF
\STATE{Update $j=j+1$;}
\UNTIL Convergence
\STATE{Obtain $\tilde{{\bf{w}}}_m^*$ and ${{\tilde{\bf{r}}}^*}$ according to (\ref{EVD}).}
\end{algorithmic}
\end{algorithm}

\emph{Remark 1:}
The proposed multiple-satellite cooperative information communication and location sensing design is essentially a multi-objective optimization problem, aiming to simultaneously minimize the location sensing accuracy and maximize the information transmission rate. In particular, in problem (\ref{OP1}), we minimize the trace of the CRB matrix while guaranteeing the quality of information communication and meeting transmit power constraints, which is known as a sensing-centric optimization. This formulation is meaningful and applicable to scenarios where location sensing tasks take priority, such as environmental monitoring, drone remote sensing, and intelligent transportation systems. However, in scenarios where information communication is the primary service goal and location sensing serves as an auxiliary function, a communication-centric optimization approach is needed. In this context, we set the information transmission rate as the objective function and the trace of the CRB matrix as the constraint, which can be formulated as
\begin{subequations}\label{OP4}
\begin{align}
\mathop {\max }\limits_{{\tilde{{\bf{w}}}_m},{\tilde{\bf{r}}}}&\
\mathop {\min} \limits_{i} {R_i}
\label{OP4a} \\
{\rm{s.t.}}
&\ \ \sum\limits_{m = 1}^M {{{\left\| {{{\bf{\Lambda }}_k}{\tilde{{\bf{w}}}_{m}}} \right\|}^2}}  + {{{\left\| {{{\bf{\Lambda }}_k}{{\tilde{\bf{r}}}}} \right\|}^2}} \le P_k^{\max }, \label{OP4b}  \\
&\ \ {\rm{ tr}}\left( {{{\left( {{{\bf{J}}^{\rm T}}\left({{{\bf{F}}_{{\bf{\Omega \Omega }}}} - {{\bf{F}}_{{\bf{\Omega }}\alpha }}{\bf{F}}_{\alpha \alpha }^{ - 1}{\bf{F}}_{{\bf{\Omega }}\alpha }^{\rm T}}\right){\bf{J}}} \right)}^{ - 1}}} \right)\leq {\widetilde{\eta}}, \label{OP4c}
\end{align}
\end{subequations}
where $\widetilde{\eta}$ denotes the required estimation accuracy for target location sensing. Then, by introducing the auxiliary variable $\Upsilon$, the original max-min objective function (\ref{OP4a}) can be converted as
\begin{subequations}\label{OP5}
\begin{align}
\mathop {\max }\limits_{{\tilde{{\bf{w}}}_m},{\tilde{\bf{r}}},\Upsilon}&\
\Upsilon \label{OP5a} \\
{\rm{s.t.}}
&\mathop {\min} {R_i} \ge \Upsilon \label{OP5b}.
\end{align}
\end{subequations}
Next, using similar convex transformation methods used to obtain problem (\ref{OP2}), the communication-centric optimization problem (\ref{OP4}) is further reformulated as
\begin{subequations}\label{OP6}
\begin{align}
\mathop {\max }\limits_{{{\bf{W}}_m},{\bf{R}},{\bf{U}},\Upsilon}&\ \Upsilon
\label{OP6a} \\
{\rm{s.t.}}
&\ {Z'_i}\left( {{{\bf{W}}_m},{\bf{R}}} \right) - {\widetilde {\rm Z}''_i}\left( {{{\bf{W}}_m},{\bf{R}}} \right) + \Upsilon \le 0, \label{OP6b} \\
&{\rm{tr}}\left( { \sum\limits_{m = 1}^M {{{\bf{\Lambda }}_k}{{\bf{W}}_m}{\bf{\Lambda }}_k^{\rm T}}+{{\bf{\Lambda }}_k}{\bf{R\Lambda }}_k^{\rm T} } \right) \le P_k^{\max }\label{OP6c} \\
&\ {\rm{tr}}\left( {{{\bf{U}}^{ - 1}}} \right)\leq \widetilde{\eta}, \label{OP6d}\\
&\ \left[ {\begin{array}{*{20}{c}}
{{{\bf{J}}^{\rm T}}{{\bf{F}}_{{\bf{\Omega \Omega }}}}{\bf{J}} - {\bf{U}}}&{{{\bf{J}}^{\rm T}}{{\bf{F}}_{{\bf{\Omega }}\alpha }}}\\
{{\bf{F}}_{{\bf{\Omega }}\alpha }^{\rm T}{\bf{J}}}&{{\bf{F}}_{\alpha \alpha }^{}}
\end{array}} \right] \succeq 0. \label{OP6e}\\
&\ {{\bf{W}}_m}\succeq0,  {\bf{R}}\succeq0, \label{OP6f} \\
&\ {\rm{Rank}}\left( {{{\bf{W}}_m}} \right) = 1,  {\rm{Rank}}\left( {\bf{R}} \right) = 1, \label{OP6g}
\end{align}
\end{subequations}
where constraints (\ref{OP6b}), (\ref{OP6f}), and (\ref{OP6g}) are obtained from (\ref{OP5b}) by using the DCA approach and SCA method, (\ref{OP6g}) denotes the transmit power limitation of LEO satellites, constraints (\ref{OP6d}) and (\ref{OP6e}) are derived from (\ref{OP4c}) according to the Schur complement theorem with auxiliary variable matrix $\bf{U}$. Finally, as with problem (\ref{OP3obj}), the standard convex communication-centric optimization problem can be solved by adding a penalty function to problem (\ref{OP6}).

In summary, although sensing-centric optimization and communication-centric optimization have different focuses and are applicable in different scenarios, the consistent formulations of performance metrics for information communication and location sensing enable both optimization problems to be effectively solved using similar convex transformation methods.

\subsection{Algorithm Analysis}
In this subsection, we provide convergence analysis and computational complexity analysis of the two mentioned algorithms in sequence.

\emph{Convergence Analysis:}
It is worth noting that the proposed target location estimation Algorithm 1 is a heuristic algorithm based on the PSO method. The convergence analysis of heuristic algorithms aims to evaluate whether the algorithm can find the optimal or a near-optimal solution within a finite number of iterations. As heuristic algorithms often rely on empirical knowledge and heuristic rules for search, their convergence analysis is typically derived from empirical and experimental studies. Thus, in order to visualize the convergence of Algorithm 1, we present in Fig. \ref{Algorithm1} the variation of the fitness value with iteration index for different numbers of particles $I_p$ in the simulation. Furthermore, for Algorithm 2, which obtains a feasible solution by iterating convex problem (\ref{OP3obj}), we have
$\mathcal{F}\big({{\bf{W}}_m^{\left( {j+1} \right)}}, {{\bf{R}}^{\left( {j+1} \right)}}\big) \leq \mathcal{F}\big({{\bf{W}}_m^{\left( {j} \right)}}, {{\bf{R}}^{\left( {j} \right)}}\big)$, where $\mathcal{F}\big({{\bf{W}}_m^{\left( {j} \right)}}, {{\bf{R}}^{\left( {j} \right)}}\big)$ denotes the objective value of problem (\ref{OP3obj}) at the $j$-th iteration. Because the power constraint (\ref{Pmax}) and the transmission rate constraint (\ref{OP1st2_2}) ensure that the trace of the CRB matrix has a lower bound, the convergence of Algorithm 2 is guaranteed according to the monotone bounded criterion \cite{Convergence analyse}. Similarly, we visualize the convergence behavior of Algorithm 2 in Fig. \ref{Algorithm2} for different LEO satellite collaboration types.

\emph{Complexity Analysis:}
For target location sensing in 3D space, the computational complexity using the grid search method is $\mathcal{O}( {n^x}{n^y}{n^z} )$, where ${n^x}$, ${n^y}$, and ${n^z}$ are related to the grid resolution on the x, y, and z axes of space, respectively. This complexity increases dramatically as the search range increases \cite{grid search}. However, the computational complexity of the PSO-based direct location sensing Algorithm 1 mainly depends on the number of particles and the maximum number of iterations. In each iteration, the algorithm evaluates fitness values for each particle and updates the particles' positions and velocities. Therefore, the complexity of the PSO-based direct location sensing algorithm is relatively low, and grows slowly when the range of interested target increases. Numerically, the computational complexity of Algorithm 1 is given by $\mathcal{O}( {N_p}{I_p} )$. Clearly, the proposed PSO-based algorithm is more efficient for direct position estimation compared to the grid search method. In addition, we note that the computational complexity of Algorithm 2 primarily arises from solving the optimization problem (\ref{OP3obj}), which is imposed by $K+M$ LMI constraints of size 1, 1 LMI constraint of size 4, and $M+1$ LMI constraints of size $NK$. Furthermore, the total number of optimization variables in problem (\ref{OP3obj}) is $z=N^2K^2M+N^2K^2+9$. Based on the computational complexity analysis in \cite{complexity analysis} for solving semi-definite programming problems by the interior-point method, the worst-case complexity for solving the problem (\ref{OP3obj}) with a desired accuracy of $\zeta>0$ can be described as $\sqrt {NKM + NK + K + M + 4}\  \delta \ln \left( {1/\zeta } \right)$, where $\delta = z( ( {N^3}{K^3}M + {N^3}{K^3}+ {N^3} + K + M + 64 ) + z( {N^2}{K^2}M + {N^2}{K^2} + {N^2} + K + M + 16 ) )$ with decision variable $z=\mathcal{O}( {N^2}{K^2}M )$ \cite{IPM}.

\section{Simulation Results}
In this section, we present the parameter settings during the numerical simulations, and show the simulation results to testify the effectiveness of the proposed algorithms. Without loss of generality, we consider a classical Walker Delta constellation similar to Starlink program, with the constellation parameters specified in Table I \cite{Walker Delta1}. We select a serving group of satellites from the considered LEO satellite constellation for simulations. In particular, we assume that all LEO satellites have the same maximum transmit power budget $P^{\max }$, and all UEs have the same required transmission rate $\eta$. If not otherwise specified, the simulation parameters are set as shown in Table II on the next page.

\begin{table}[ht]
\small
\centering
\caption{Parameters Of LEO Satellite Constellation }\label{Simulation}
\begin{tabular}{|c|c|}
\hline
Parameter & Value  \\ \hline
Orbital altitude & $H^{\#}=550$ km \\\hline
Number of orbital planes & $P^{\#}=72$ \\\hline
Number of LEO satellites per orbital plane & $N^{\#}=22$ \\\hline
Orbital inclination & $I^{\#}=53^{\circ}$ \\\hline
Phase factor & $F^{\#}=1$ \\\hline
\end{tabular}
\end{table}

\begin{table*}[ht]
\small
\centering
\caption{Simulation Parameters }\label{Simulation}
\begin{tabular}{|c|c|}
\hline
Parameter & Value \\ \hline
Number of UPA antennas & $N=N_x\times N_z=4\times 4$ \\\hline
Number of LEO satellites in a group & $K=5$ \\\hline
Number of UEs covered & $M=10$ \\\hline
Rician factor & $\lambda _k=10$ dB \\\hline
Speed of light & $c=3\times 10^{8}$ m/s \\\hline
Signal frequency & $f=35$ GHz \\\hline
Boltzmann constant & $\kappa=1.38 \times 10^{-23}$ J/m \\\hline
Channel bandwidth & $B=20$ MHz \\\hline
Ratio of receive antenna gain to noise temperature & $G_m/T=34$ dB/K \\\hline
Distance between LEO satellites and UEs & $d_{k,m}\in[550\sim2700]$ km \\\hline
Distribution of rain attenuation gain & $\mu _r=-2.6$ dB, $\sigma _r^2=1.63$ dB \\\hline
Maximum antenna gain of LEO satellites & ${b_{k}^{\max}}=16$ dBi \\\hline
3-dB angle & $\varepsilon_k^{3dB}=0.4^{\circ}$  \\\hline
Noise power & $\sigma _i^2=\sigma_n ^2=-110$ dBm \\\hline
Adjacent antenna spacing related to signal wavelength & $d=\frac{1}{2}\lambda$ \\\hline
Number of particles & $I_p=50$  \\\hline
Maximum number of iterations & $N_p=40$  \\\hline
Individual and swarm learning factor & $c_1=1.5$, $c_2=1.5$   \\\hline
Maximum and minimum inertia weight & $\varpi_{\max}=0.8$, $\varpi_{\max}=0.4$ \\\hline
Maximum transmit power budget of LEO satellites & $P_k^{\max }=P^{\max }=30$ dBm \\\hline
Required transmission rate of UEs & $\eta _i=\eta=2$ bps/Hz \\\hline
Initial penalty factor & $\rho=10$ \\\hline
Amplification coefficient & $\iota=1.5$ \\\hline
Penalty accuracy & $\Delta=10^{-4}$ \\\hline
\end{tabular}
\end{table*}

\begin{figure}
 \centering
\includegraphics [width=0.47\textwidth] {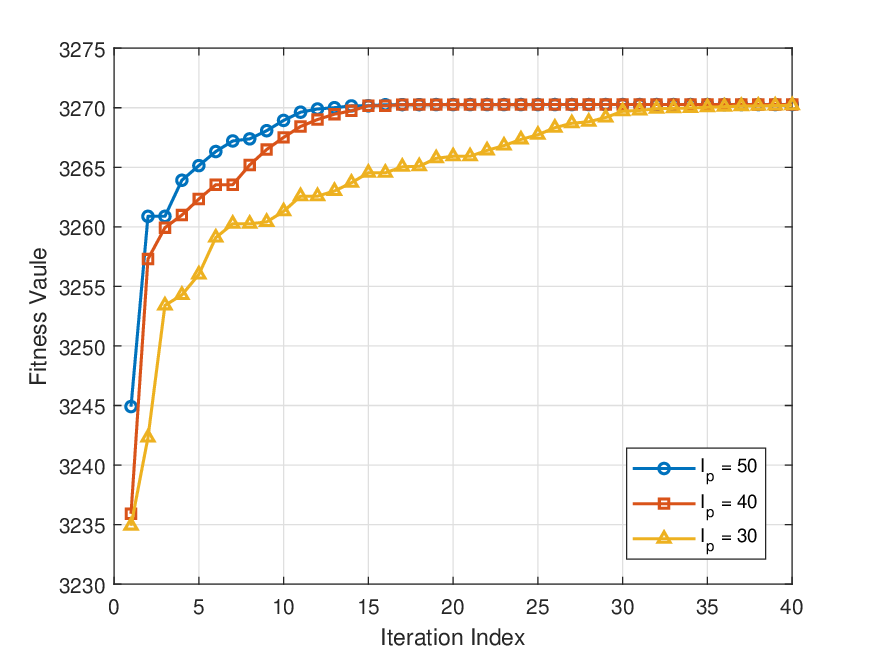}
\caption {Convergence behavior of Algorithm 1.}
\label{Algorithm1}
\end{figure}
Firstly, we present the convergence behavior of Algorithm 1 for different numbers of particles $I_p$. In Fig. \ref{Algorithm1}, we utilize the fitness value of the historical optimal location of the particle swarm, denoted as $F\left({{{\bf{g}}^{{\rm{best}}}}}\right)$, to assess the convergence performance. Typically, the fitness value in PSO algorithms represents the quality or performance evaluation of a solution in the problem space. Higher fitness values indicate a better fit between the particle's position solution and the observed data, reflecting a higher likelihood. It can be seen that for the heuristic Algorithm 1, when a larger number of particles $I_p$ are employed, the global search capability of the algorithm is enhanced, allowing it to find the optimal value within a smaller number of iterations. However, it should be noted that increasing the number of particles $I_p$ leads to an increase in computing resource consumption. Specifically, it results in higher computational complexity and increased memory usage. Each particle needs to store and update information such as location, velocity, and fitness value. Therefore, when applying the Algorithm 1 in practice, it is necessary to select an appropriate number of particles $I_p$ and maximum number of iterations $N_p$ based on the available computing resources and the location search scale.

\begin{figure}
 \centering
\includegraphics [width=0.47\textwidth] {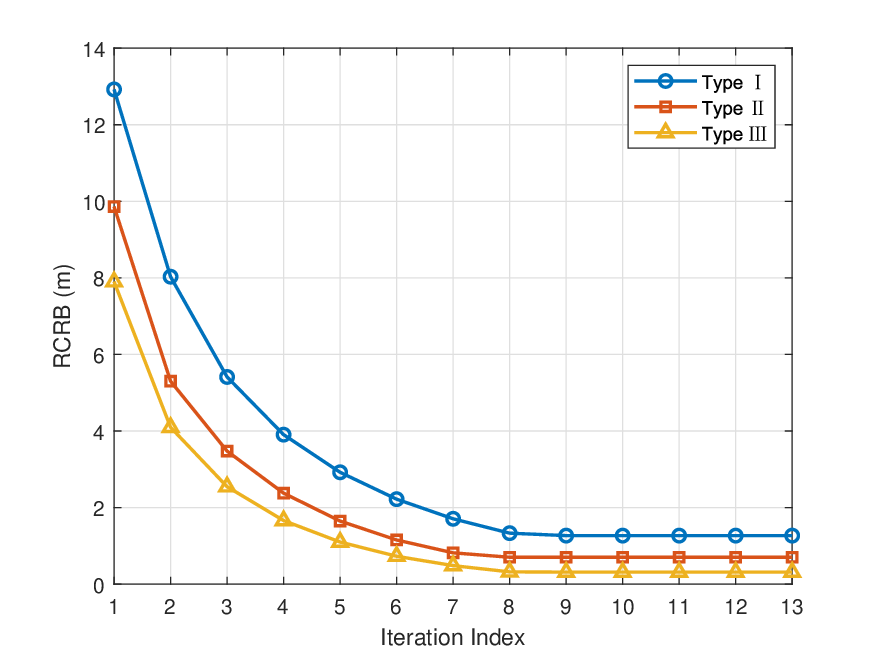}
\caption {Convergence behavior of Algorithm 2.}
\label{Algorithm2}
\end{figure}

\begin{figure}
 \centering
\includegraphics [width=0.47\textwidth] {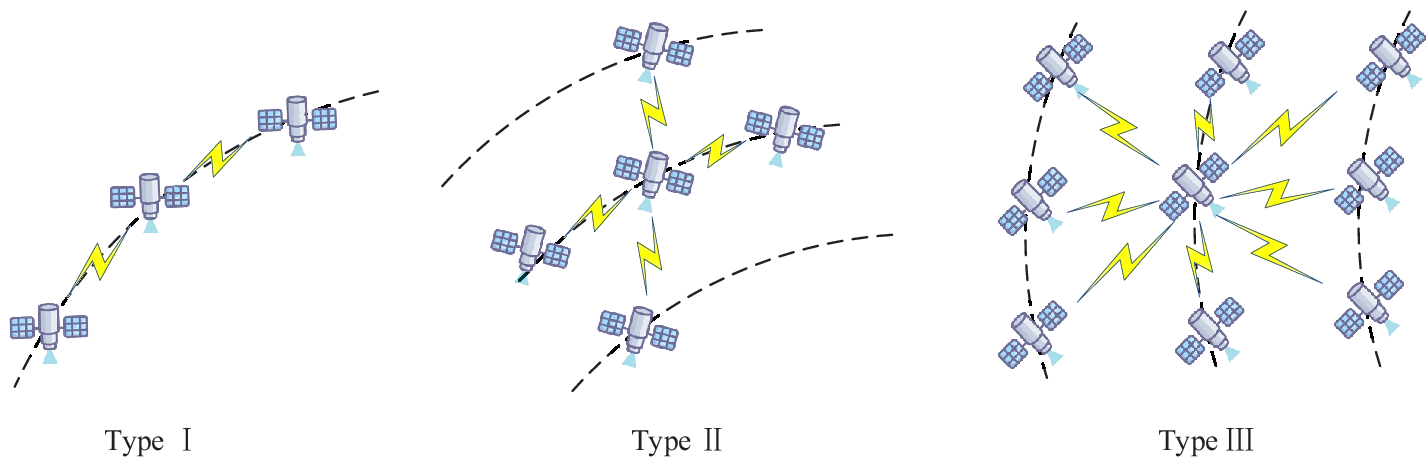}
\caption {Types of LEO satellites collaboration.}
\label{Fig_type}
\end{figure}
Secondly, Fig. \ref{Algorithm2} illustrates the convergence behavior of Algorithm 2 for different LEO satellite collaboration types. In particular, collaboration type \uppercase\expandafter{\romannumeral 1} indicates that the central satellite establishes inter-satellite links only with the adjacent satellites on the same orbital plane. Collaboration type \uppercase\expandafter{\romannumeral 2} indicates that the central satellite establishes inter-satellite links with the adjacent satellites on the same orbital plane and the nearest satellite on each adjacent orbital planet. Finally, collaboration type \uppercase\expandafter{\romannumeral 3} indicates an extension of type \uppercase\expandafter{\romannumeral 2}, where the central satellite establishes inter-satellite links with additional satellites on adjacent orbital planes for joint information communication and location sensing. These types of LEO satellites collaboration are visualized in Fig. \ref{Fig_type}, and the remaining simulations in this section are conducted for the type II case. As can be seen from Fig. \ref{Algorithm2}, the root of CRB (RCRB) value monotonically decreases with iterations, eventually converging to a stable point within 10 iterations for different LEO satellite collaboration types. This observation indicates that the computational complexity of the proposed Algorithm 2 is affordable for real-world applications of LEO satellite constellations, and the location sensing performance enhances with the growing number of collaborative satellites.

\begin{figure}
 \centering
\includegraphics [width=0.47\textwidth] {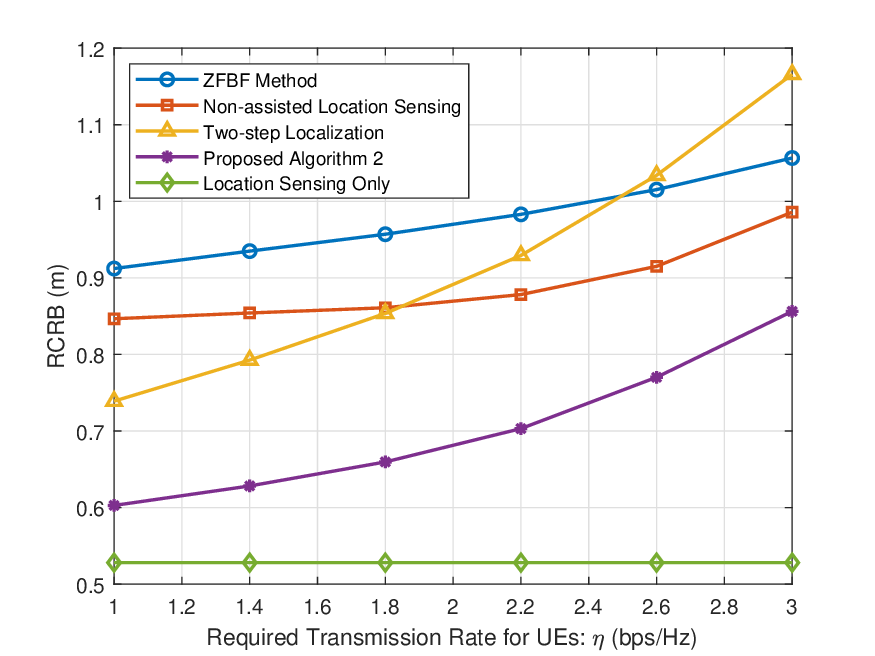}
\caption {Performance comparison for different multiple-satellite cooperative algorithms.}
\label{compare}
\end{figure}
Next, in Fig. \ref{compare} we investigate the enhancement of the proposed Algorithm 2 compared to other localization or beamforming design methods, where ``ZFBF Method" refers to the design of communication transmit beamforming by utilizing the zero-forcing beamforming method \cite{ZFBF Method}, ``Non-assisted Location Sensing" refers to a conventional integrated communication and sensing approach in which the communication signals are not utilized to assist location sensing \cite{Non-assisted Location Sensing}, ``Two-step Localization" refers to the classic two-step passive localization technology \cite{Two-step Localization}, and ``Location Sensing Only" refers to an optimization design that disregards the transmission rate constraint (\ref{OP1st2}) to demonstrate an upper bound on the performance of location sensing. It is evident that the proposed Algorithm 2 exhibits a superior performance. Meanwhile, a performance trade-off between information communication and location sensing can also be identified, i.e, the location sensing accuracy deteriorates significantly as the required communication rate increases. The gap between the lines ``Proposed Algorithm 2" and ``Location Sensing Only" describes the loss of location sensing accuracy due to the provision of information communication services. Therefore, it is meaningful to balance the performance of information communication and location sensing according to the actual circumstances.

\begin{figure}
 \centering
\includegraphics [width=0.47\textwidth] {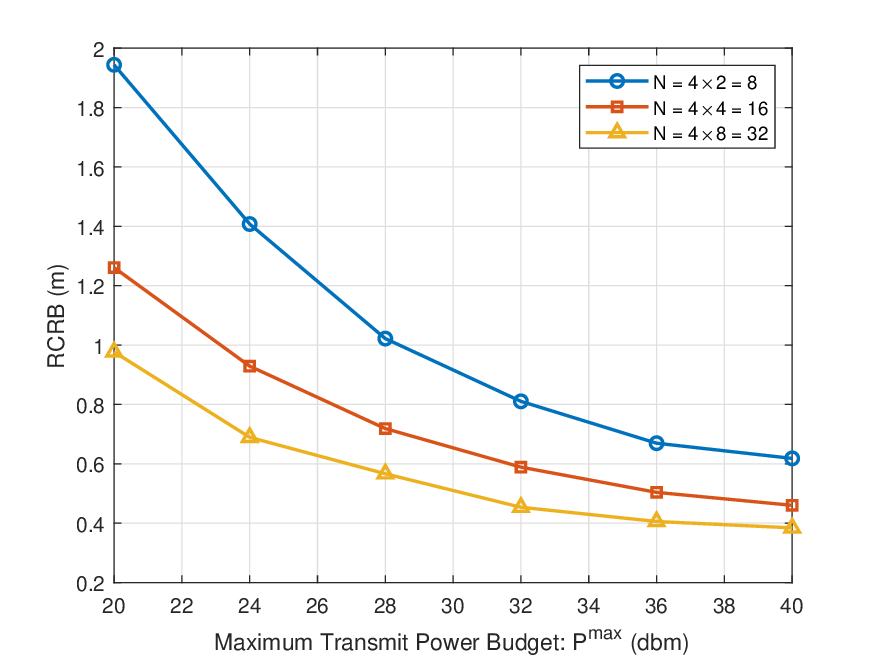}
\caption {RCRB versus the maximum transmit power budget for different scales of UPA at LEO satellites.}
\label{Pmax_N}
\end{figure}
Fig. \ref{Pmax_N} examines the impact of different scales of UPA at LEO satellites and maximum satellite transmit power budgets on the system performance. It is seen that RCRB decreases as the maximum transmit power budget $P^{\max}$ increases. This is due to the fact that dual-function signals experience significant attenuation in the satellite-terrestrial channels. By increasing the signal transmit power, higher signal-to-noise ratio and signal measurement accuracy can be achieved at the receiver, thereby simultaneously improving the quality of information communication and the accuracy of location sensing. In addition, the value of RCRB exhibits a decreasing trend as the scale of the UPA at the LEO satellites increases. This is because a larger scale of UPA enhances the efficiency of information transmission and provides a larger number of spatial degrees of freedom for location sensing. Consequently, it becomes essential to deploy a UPA with appropriate scale at the LEO satellites in practical applications in order to strike a balance between construction costs and system performance.

\begin{figure}
 \centering
\includegraphics [width=0.47\textwidth] {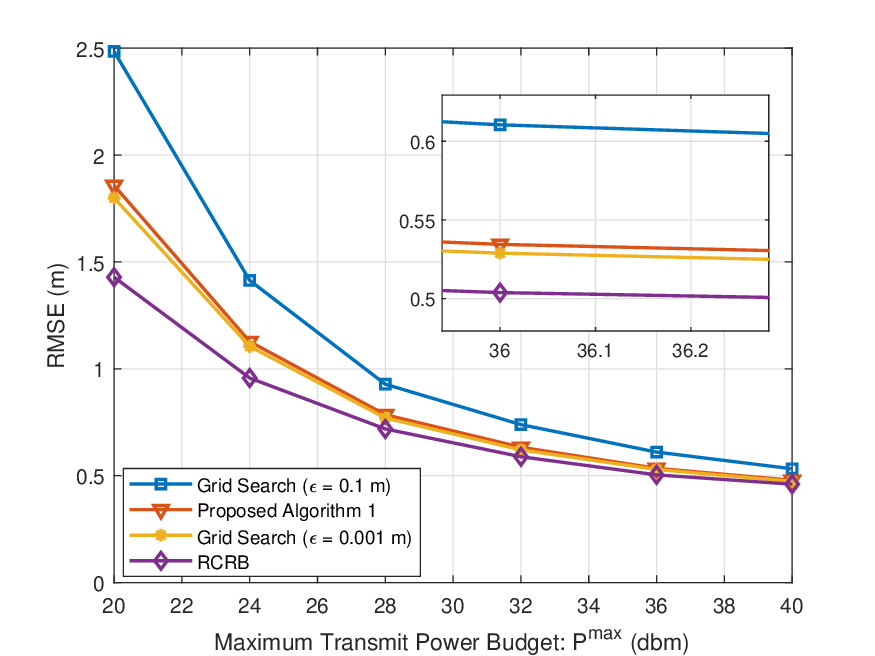}
\caption {Performance comparison with the baseline algorithms.}
\label{compareR1}
\end{figure}

\begin{figure}
 \centering
\includegraphics [width=0.47\textwidth] {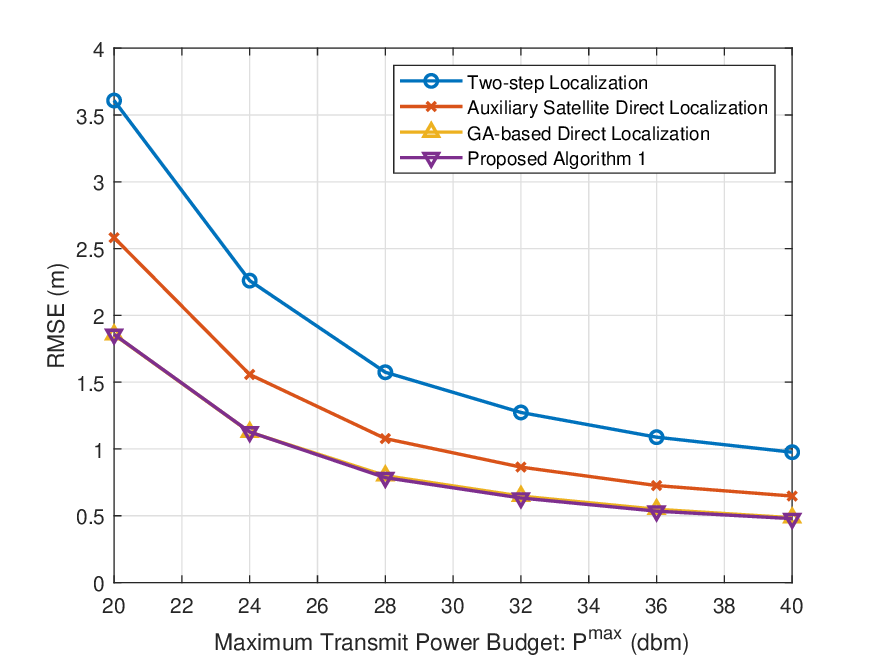}
\caption {RMSE versus the maximum transmit power budget for different location sensing algorithms.}
\label{compareR2}
\end{figure}
We have conducted $1000$ Monte Carlo simulation runs to examine the root of mean square error (RMSE) between the estimated target location obtained through Algorithm 1 and the true location, i.e. ${\rm{RMSE}} = \sqrt {\frac{1}{{1000}}\sum\limits_{i = 1}^{1000} {{{\left\| {{{{\bf{\widehat p}}}_i} - {{\bf{p}}_i}} \right\|}^2}} } $. In Fig. \ref{compareR1} and Fig. \ref{compareR2}, we focus on the RMSE obtained from the Monte Carlo simulations for the proposed Algorithm 1, the Two-step Localization, the Genetic Algorithm (GA)-based Direct Localization, the Auxiliary Satellite Direct Localization and the Grid Search method \cite{grid search} with search accuracy of $\epsilon$, respectively. Among them, ``Auxiliary Satellite Direct Localization" refers to performing direct position estimation at each auxiliary satellite and then averaging the target estimation results from multiple auxiliary satellites with weighted averaging at the central satellite for target localization. Moreover, ``GA-based direct localization" replaces PSO with GA in Algorithm 1 to tackle the non-convex optimization problem of MLE. At the same time, we consider the RCRB obtained by Algorithm 2 which serves as theoretical lower bound for the actual location sensing performance metric RMSE. It can be observed that due to the influence of noise, the RMSE of Algorithm 1 is slightly larger than the optimized RCRB. However, as the maximum transmit power budget $P^{\max}$ increases, leading to an increase in signal-to-noise ratio, the values of RCRB and RMSE of Algorithm 1 and the gap between them gradually decrease, which confirms the reference of using the CRB as a metric for evaluating location sensing performance. In addition, the performance of the proposed Algorithm 1 is close to the Grid Search with a search accuracy of 0.001 m and significantly better than traditional Two-step Localization and Auxiliary Satellite Direct Localization. According to the complexity analysis in Section \uppercase\expandafter{\romannumeral 3}. C, the computational complexity of the PSO-based Algorithm 1 is much lower than the high-accuracy grid search method, which fully demonstrates the excellent performance of the proposed Algorithm 1. Furthermore, Algorithm 1 and GA-based direct localization are both relatively simple localization methods and exhibit similar performance. However, the parameter tuning process for GA-based direct localization is more complex, so we finally adopt the PSO-based Algorithm 1. Further, noting that the two-step localization method reduces inter-satellite information exchange by first obtaining quantized localization parameters at each satellite before uploading them to the central satellite. On the contrary, the proposed PSO-based direct location sensing Algorithm 1 achieves higher localization accuracy at the expense of transmitting more sensing signal data containing the target's position information among the satellites. Fortunately, the use of inter-satellite optical links, which employ laser communication technology to provide high-speed signal transmission, makes Algorithm 1 feasible and practical, even with frequent inter-satellite signal exchange.

\begin{figure}
 \centering
\includegraphics [width=0.47\textwidth] {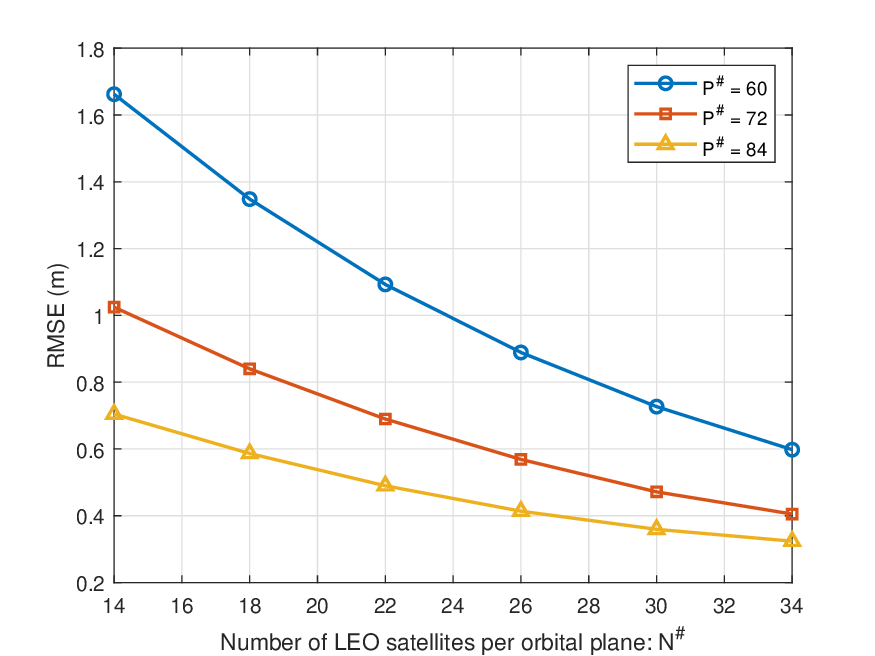}
\caption {RMSE versus the number of LEO satellites per orbital plane for different numbers of orbital planes.}
\label{satellite_number}
\end{figure}
Lastly, we show the location sensing accuracy of the proposed Algorithm 1 for different LEO satellite constellation parameters through Monte Carlo simulations in Fig. \ref{satellite_number}. It is evident that as the number of orbital planes and the number of LEO satellites per orbit increase, the RMSE gradually decreases. Since the intensive deployment of LEO satellites significantly reduces the distances between LEO satellites and interested target as well as between LEO satellites and UEs, the channel conditions for information communication and location sensing are improved. This also explains why the new generation LEO satellite constellation projects plan to deploy more satellites to provide seamless services simultaneously. Nevertheless, it is worth noting that the benefits of the dense deployment of LEO satellites diminish as the number of LEO satellites increases. Additionally, this can lead to risks of space collisions, excessive signal interference, and increased costs and resource consumption. Consequently, it is crucial to carefully determine the parameters for LEO satellite constellations based on service requirements and the need for continuous coverage, in order to achieve a balance between performance and system feasibility as well as cost-effectiveness in practical applications.

\section{Conclusion}
This paper presented a dual-function LEO satellite constellation framework for realizing multiple-satellite collaborative information communication and location sensing simultaneously. To improve the information transmission rate and location sensing accuracy, a joint communication beamforming and sensing waveform design was proposed and formulated as a complex non-convex optimization problem. To solve this problem, a penalty function-based iterative optimization algorithm was put forward to obtain a feasible solution. Simulation results with various parameters demonstrated that through the proposed PSO-based direct location sensing Algorithm 1 and multiple-satellite cooperative information communication and location sensing Algorithm 2, the information communication and location sensing can achieve superior performance to traditional methods at a relatively low complexity level simultaneously. In future work, it would be interested in exploring more functional integration and cross-orbit satellite cooperation.

\begin{appendices}
\section{Calculation of Jacobian matrixs ${{\bf{J}}}$}
The Jacobian matrix ${{\bf{J}}}$ is obtained by the partial derivation of the angle parameter $\bf{\Omega}$ with respect to the location parameter $\bf{p}$, i.e.
\begin{equation}\label{J}
{{\bf{J}}} = \frac{{\partial {\bf{\Omega }}}}{{\partial {\bf{p}}}} = \left[ {\frac{{\partial {\bf{\Omega }}}}{{\partial {p^x}}},\frac{{\partial {\bf{\Omega }}}}{{\partial {p^y}}},\frac{{\partial {\bf{\Omega }}}}{{\partial {p^z}}}} \right],
\end{equation}
where
$\frac{{\partial {\bf{\Omega }}}}{{\partial {p^x}}} = \big[ \frac{{\partial {\theta _1}}}{{\partial {p^x}}}; \cdots ;\frac{{\partial {\theta _k}}}{{\partial {p^x}}}; \cdots ;\frac{{\partial {\theta _K}}}{{\partial {p^x}}};\frac{{\partial {\varphi _1}}}{{\partial {p^x}}}; \cdots ;\frac{{\partial {\varphi _k}}}{{\partial {p^x}}}; \cdots$ $ ;\frac{{\partial {\varphi _K}}}{{\partial {p^x}}} \big]$, and ${\frac{{\partial {\bf{\Omega }}}}{{\partial {p^y}}}}$ and ${\frac{{\partial {\bf{\Omega }}}}{{\partial {p^z}}}}$ have similar forms. Further, according to (\ref{theta}), we take the partial derivative to obtain
\begin{equation}\label{J_1}
\frac{{\partial {\theta _k}}}{{\partial {p^x}}} = \frac{{\left( {{p^x} - q_k^x} \right)\left( {{p^z} - q_k^z} \right)}}{{D_k^2\sqrt {{{\left( {{p^x} - q_k^x} \right)}^2} + {{\left( {{p^y} - q_k^y} \right)}^2}} }},
\end{equation}
\begin{equation}
\frac{{\partial {\theta _k}}}{{\partial {p^y}}} = \frac{{\left( {{p^y} - q_k^y} \right)\left( {{p^z} - q_k^z} \right)}}{{D_k^2\sqrt {{{\left( {{p^x} - q_k^x} \right)}^2} + {{\left( {{p^y} - q_k^y} \right)}^2}} }}
\end{equation}
and
\begin{equation}
\frac{{\partial {\theta _k}}}{{\partial {p^z}}} =  - \frac{{\sqrt {{{\left( {{p^x} - q_k^x} \right)}^2} + {{\left( {{p^y} - q_k^y} \right)}^2}} }}{{D_k^2}},
\end{equation}
where ${D_k} = \sqrt {{{\left( {{p^x} - q_k^x} \right)}^2} + {{\left( {{p^y} - q_k^y} \right)}^2} + {{\left( {{p^z} - q_k^z} \right)}^2}} $. According to (\ref{varphi}), we have
\begin{equation}
\frac{{\partial {\varphi _k}}}{{\partial {p^x}}} = \frac{{q_k^y - {p^y}}}{{{{\left( {{p^x} - q_k^x} \right)}^2} + {{\left( {{p^y} - q_k^y} \right)}^2}}},
\end{equation}
\begin{equation}
\frac{{\partial {\varphi _k}}}{{\partial {p^y}}} = \frac{{{p^x} - q_k^x}}{{{{\left( {{p^x} - q_k^x} \right)}^2} + {{\left( {{p^y} - q_k^y} \right)}^2}}}
\end{equation}
and
\begin{equation}\label{J_2}
\frac{{\partial {\varphi _k}}}{{\partial {p^z}}} = 0.
\end{equation}
Finally, by substituting equations (\ref{J_1})-(\ref{J_2}) into (\ref{J}), the calculation of Jacobian matrix ${{\bf{J}}}$ is finished.

\end{appendices}

\end{document}